\documentclass[usenatbib,useAMS]{mn2e}
\usepackage{graphicx}

\title[The colour-magnitude relations of ClJ1226.9+3332]{The colour-magnitude relations of ClJ1226.9+3332, a massive cluster of galaxies at $z=0.89$}

\author[S.C. Ellis et al.]
{S.C.~Ellis$^{1}$\thanks{E-mail: sce@aao.gov.au}, L.R.~Jones$^{2}$, D.~Donovan$^{3}$, H.~Ebeling$^{3}$ and H.G.~Khosroshahi$^{2}$\\
$^{1}$Anglo-Australian Observatory, PO Box 296, Epping, NSW 2121, Australia.\\  
$^{2}$School of Physics and Astronomy, University of Birmingham, Birmingham, B15 2TT, UK. \\
$^{3}$IfA, University of Hawaii, 2680 Woodlawn Dr., Honolulu, HI 96822, USA.}

\date{Accepted .....................; Received .....................;
in original form .......................}

\def\gtrsim{\mathrel{\hbox{\rlap{\hbox{\lower4pt\hbox{$\sim$}}}\hbox{$>$}}}}
\def\lsim{\mathrel{\hbox{\rlap{\hbox{\lower4pt\hbox{$\sim$}}}\hbox{$<$}}}}

\begin{document} 

\maketitle

\begin{abstract}
The colour-magnitude relations of one of the most massive ($\approx 10^{15}$M$_{\odot}$), high redshift ($z=0.89$) clusters of galaxies known have been studied.  Photometry has been measured in the \emph{V}, \emph{R}, \emph{I}, \emph{z}, \emph{F606W}, \emph{F814W}, \emph{J} and \emph{K} bands to a depth of $K\approx 20.5=K^{*}+2.5$ and spectroscopy confirms 27 \emph{K} band selected galaxies as members of the cluster.  The $V-K$ colours are equivalent to a rest-frame colour of $\approx2700{\rm \AA}\ -J$, and provide a very sensitive measure of star-formation activity.  Hubble Space Telescope imaging with the Advanced Camera for Surveys has been used to morphologically classify the galaxies.

The cluster has a low early-type fraction compared to nearby clusters, with only 33 per cent of the confirmed cluster members having types E or S0.

The early-type member galaxies form a clear red-sequence in all colours.  The scatter and slope of the relations show no evolution compared to the equivalent Coma cluster relations, suggesting the stellar populations are already very old at $z=0.89$.  The normalisation of the relations has been compared to models based on synthetic stellar populations, and are most consistent with stellar populations forming at $z_{{\rm f}} \gtrsim 3$.

Some galaxies of late-type morphology were found to lie on the red-sequence of the colour-magnitude relation, suggesting that they have very similar stellar populations to the early-type galaxies.  

These results present a picture of a cluster in which the early-type galaxies are all old, but in which there must be future morphological transformation of galaxies to match the early-type fraction of nearby clusters.  In order to preserve the tight colour-magnitude relation of early-types seen in nearby clusters, the late-type galaxies must transform their colours, through the cessation of star-formation, before the morphological transformation occurs.  Such evolution is observed in the late-types lying on the colour-magnitude relation.

\end{abstract}


\section{Introduction}

Observations of galaxies in clusters reveal several different evolutionary trends.
On the one hand there is an apparent change in the galaxy populations of clusters with redshift.  Whilst the cores of nearby clusters are dominated by early-type elliptical (E) and lenticular (S0) galaxies (\citealt{dre80}), at high redshift there exists a significant population of blue, actively star-forming and post-starburst galaxies which are almost absent at low redshifts (\citealt{but84}).  The higher blue fraction of galaxies at high redshift may indicate that the morphological composition of the clusters is evolving, with high redshift clusters containing a larger fraction of spiral galaxies.    Indeed,  \citet{dre97} show that there is a 
decrease in the fraction of S0s with increasing redshift, and a corresponding increase in the fraction of spirals.  These observations seem to require some transformation of spirals to S0s, coinciding with a decrease in star-formation rate (e.g.\ \citealt{lar80}), possibly arising from the infall of field spirals into the cluster and the subsequent quenching of the star-formation.  

On the other hand observations of early-type galaxies in clusters out to redshifts $z\gtrsim 1$ reveal consistently old stellar populations, with correspondingly early assembly epochs. In particular, the tight correlation between the colour and magnitudes of cluster galaxies at both low (e.g.\ \citealt{bow98}) and high (e.g.\ \citealt{sta98}; \citealt{hol04}) redshifts, the evolution of the fundamental plane (e.g.\ \citealt{van98}) and the evolution of the $K$ band luminosity function (e.g.\ \citealt{tre98b}; \citealt{dep99};  \citealt{nak01}; \citealt{nel01}; \citealt{kod03}; \citealt{tof03}; \citealt{ell04}; \citealt{tof04}; \citealt{stra06}) all reveal properties consistent with formation at early epochs ($z \gtrsim 2$), although LFs derived at bluer wavelengths shows a small fraction of the stellar mass may be formed at lower redshifts (e.g.\ \citealt{nel01}; \citealt{and04}).  

Note also, that there are some problems with the interpretation of the Butcher-Oemler effect as an indication of morphological transformation.  \citet{and97} show that the luminosity function of S0 galaxies in the $z=0.023$ Coma cluster is consistent with the S0 LF in the $z=0.41$ cluster, which is at odds with an evolving population of S0s.
\citet{dep03b} find very little increase in the blue fraction of galaxies when selection is made in the near-infrared, suggesting that the blue, high redshift galaxies may be low mass systems rather than massive spirals. \citet{burs05} show that S0s are brighter, and by implication more massive, on average than spirals in the NIR, which is inconsistent with a formation resulting from the stripping of mass from spiral galaxies.  


A further complication to the above picture is that numerical simulations of the hierarchical formation of clusters of galaxies reveal that as many as 50\% of present day cluster galaxies may not have been within the cluster at $z=1$, having since fallen into the cluster as part of the assembly process (\citealt{del04}).  Thus any infalling galaxies which are progenitors for today's early-type galaxies must evolve to have remarkably similar properties to the galaxies already in the cluster core, to account for the extremely small scatter of the local CMR of early-type galaxies.

These issues have been studied in the CMRs of one of the most massive ($M=1.4\pm 0.2 \times 10^{15}$ M$_{\odot}$) cluster of galaxies known at $z>0.6$ ($z=0.89$).  
It thus provides an excellent opportunity to study the properties of galaxies in high-density environments at high redshifts.  Galaxies were selected in the near-infrared, providing a robust measurement of the luminosities of the old stellar populations,  which is closely related to the dynamical mass of the galaxies (\citealt{gav96}).   Spectroscopy provides accurate determination of cluster membership for the brighter galaxies, and Hubble Space Telescope (HST) imaging with the Advanced Camera for Surveys (ACS) provides morphological classifications.  We have obtained near-infrared and optical photometry (\emph{V}, \emph{R}, \emph{I}, \emph{z}, \emph{F606W}, \emph{F814W}, \emph{J} and \emph{K}) from which CMRs have been constructed.  The evolution of the galaxy population has been investigated via 
their scatter, normalisation and slope.


The paper is organised as follows.  In section~\ref{sec:cluster} we describe the cluster, and the sample. 
Section~\ref{sec:photo} describes the photometry.  Section~\ref{sec:morph} details the morphological classification of the galaxies.  In Section~\ref{sec:earlyfrac} we discuss the fraction of early-type galaxies in the cluster.  Section~\ref{sec:cmr} describes the calculation of the CMR and presents the results.  Section~\ref{sec:evol} investigates the evolution of the CMR via its scatter, normalisation and slope.  Finally a discussion of the results and their implications for galaxy evolution in clusters are presented in section~\ref{sec:discuss}.
A cosmology of $H_{0}=70$km s$^{-1}$ Mpc$^{-1}$, $\Omega_{{\rm M}}=0.3$ and $\Omega_{\Lambda}=0.7$ has been assumed throughout.

\section{ClJ1226.9+3332}
\label{sec:cluster}

\begin{figure}
\includegraphics[width=1\columnwidth]{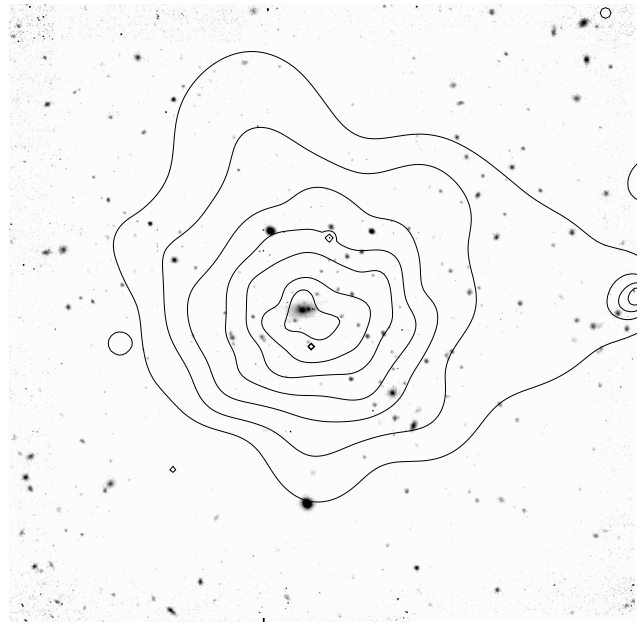}
\caption{UKIRT UFTI \emph{K} band image of the $z=0.89$ cluster
  ClJ1226 with adaptively smoothed XMM X-ray contours from
  \protect\citealt{mau04}. The image is $\sim 2.9$ arcmin square.}
\label{fig:1226}
\end{figure}

ClJ1226.9+3332, hereafter ClJ1226, (\protect\citealt{ebe01};
\protect\citealt{cag01}; \protect\citealt{mau04}) is at a redshift $z=0.892 \pm 0.007$.  It has a bolometric
X-ray luminosity of $L_{\mathrm{X}}=5.3^{+0.2}_{-0.2} \times 10^{45}$
ergs s$^{-1}$, an X-ray temperature of
$T_{\mathrm{X}}=11.5^{+1.1}_{-0.9}$ keV and a total mass
$M_{\mathrm{total}}=1.4^{+0.2}_{-0.2} \times 10^{15}$ M$_{\odot}$
(\protect\citealt{mau04}), making it one of the most massive clusters
known at high redshifts. The cluster appears remarkably relaxed in
X-rays indicating little dynamic activity, although recent XMM observations reveal temperature substructure and entropy profiles which suggest that the cluster may be undergoing a line-of-sight merger (Ben Maughan, private communication).  The \emph{K} band image with
adaptively smoothed  \emph{XMM} X-ray contours (from \citealt{mau04}) is shown in Figure~\ref{fig:1226}.

\subsection{Observations}

We have obtained near-infrared imaging and photometry of ClJ1226, from which the initial galaxy selection was made, and a study of the \emph{K} band galaxy luminosity function was made (\citealt{ell04}).  This was followed up with optical imaging and photometry, including high spatial resolution HST imaging from which morphological classifications were made, as well as multi-object spectroscopy.  These observations will now be described in turn.

\subsubsection{Near-Infrared imaging and photometry}

Observations were made on the 18th, 19th and 20th April 2001 at the 3.8m United Kingdom Infrared Telescope ({\sc UKIRT}) on Mauna Kea, Hawaii using the UKIRT Fast Track Imager ({\sc UFTI}, \citealt{roc02}) camera, a $1024 \times 1024$ pixel HgCdTe array, with a pixel size of 0.091 arcseconds.  Observations were made with the K98 and J98 filters (50\% cut-offs  2.03--2.37$\mu$m and 1.17--1.33$\mu$m respectively) with an exposure time of 66s per frame and the seeing was
typically $\approx 0.5$ arcsec in \emph{K}.  The total exposure times were 23946s in \emph{K} and 26738s in \emph{J}.  The photometry was complete to $K\approx 20.5$ mag and $J\approx21$ mag, determined by the magnitude at which the number counts of galaxies begins to fall. A spatial dithering of 15 arcsec was employed and a mosaic was constructed from four pointings, since the field of view of 90x90 arcsec covered only a fraction of the virial radius. 
On each night standard stars (from \protect\citealt{haw01}) were observed in order to calibrate the zero-point, air mass coefficient, and colour coefficient.  One night had patchy cloud cover, and thus yielded inaccurate photometric calibration.  This night was deemed to be non-photometric,
and a calibration was obtained via observations of the same fields on photometric nights.
Non-photometric data were then weighted according to their shorter effective exposure times.
The observations covered an area of $2'47'' \times 3'1''$, corresponding to a physical size of $1.3 \times 1.4$ Mpc. 

\subsubsection{Optical imaging and photometry}

We have obtained \emph{V}, \emph{R}, \emph{I} and \emph{z} band imaging of ClJ1226 at the 8.2m Subaru Telescope on Mauna Kea, Hawaii, using SuprimeCam, a prime focus camera consisting of 10 MIT/LL 2048$\times$4096 pixel arrays, with 0.2 arcseconds per pixel, yielding a total field of view of 34$\times$27 arcminutes.

The total exposure times were 2160s in \emph{V}, 2880s in \emph{R}, 1920s in \emph{I}  and 1080s in \emph{z}, yielding completeness limits of 24, 24, 24, 23 mag respectively.  Standard stars from \citet{lan92} were observed on each night.  

\subsubsection{Multi-object spectroscopy}

In order to assess cluster membership, spectra were obtained and redshifts measured using the 10m Keck Telescope on Mauna Kea, with the Low Resolution Imaging Spectrometer (LRIS), and later using the 8.1m Gemini North Telescope, also on Mauna Kea, with the Gemini Multi-object spectrograph (GMOS).


Forty-five galaxies have been confirmed as cluster members, 27 of which are within the UFTI \emph{K} band field of view.  The spectroscopic completeness of the \emph{K} band selected galaxies is 82 per cent at $K<18$ and 61 per cent at $K<19$.

\subsubsection{HST imaging}

As part of a larger project to study the lensing and morphological properties of the galaxies in ClJ1226 observations have been obtained with the 2.4m HST using the ACS.  The ACS was used with the Wide Field Camera (WFC), which consists of two 2048 $\times$ 4096 pixel Scientific Imaging Technology CCDs, with a field of view of 202 $\times$ 202 arcseconds.

Six tiles of 4000s exposures each were mosaiced, in both the \emph{F606W} and \emph{F814W} filters, giving completeness limits of 25 and 23 mag respectively.








\section{Photometry.}
\label{sec:photo}
\label{photometry}

\subsection{Standard stars.}
\label{standards}

The photometry of the galaxies was calibrated using standard stars 
observed on the same nights as the cluster fields.   The true magnitude of any celestial body is
taken to be
\begin{equation}
\label{eqn:mag}
m_{{\rm true}} = ZP - 2.5 {\rm log} (CR) + A{\rm sec}z + B(J-K)_{{\rm true}}
\end{equation}

where $ZP$ is the
zero-point magnitude, $CR$ is the count rate, $A$ is the coefficient of extinction per unit airmass,
$z$ is the zenith angle (sec$z$ therefore being the airmass) and $B$ is the
colour coefficient necessary due to differences in the response of the combination of
camera and filters used and the standard star system. 
Galactic extinction (0.06 mag in \emph{V}, 0.007 mag in \emph{K}), was assumed to be negligible given the possible systematic uncertainties of $\sim 0.1$ mag for each band.

\subsubsection{NIR, UKIRT}
Values of $ZP$, $A$ and $B$ were determined from observations of standard stars 
taken on each night.  Standard stars were selected from the UKIRT faint standards list (\protect\citealt{haw01}).
There were typically  6 stars of varying colours  observed each night between 1 and 3 times, each at differing airmasses. 

Whilst the value of $A$ and $ZP$ may vary from night to night, the value of the 
colour coefficient 
$B$ should be almost constant. Thus for the NIR a single value of $B$ was measured from all 
the photometric nights
combined. The value we obtained is consistent with zero, as expected from the filter design (\protect\citealt{tok02}).

 An independent value of $ZP$ was measured
 for each night, and for the two photometric nights,
an independent value of $A$ was measured, and checked for consistency. Where this was not possible, on the cloudy night, a value of
 $A$   
from other nights was used and checked for consistency.  All were found to be consistent, and a single value from one the photometric nights was used subsequently.
In any case the low extinction in the \emph{K} band and the low airmasses at which most
 observations
were performed combine to make this a relatively small correction ($<$0.2 mag).

Note that the final photometric error, as estimated from the scatter
in the standard star measurements, is less than 0.1 mag.
The photometric calibrations are summarised in Table~\ref{tab:cal}.

\subsubsection{Optical, Subaru}

For the optical SuprimeCam data standard star fields from \citet{lan92} were observed each night.  There were typically two fields per night per filter, viz.\ SA101, SA98 and SA95 for \emph{I}, SA95 and SA92 for \emph{R} and SA113 for \emph{V}.  Note that each one of these fields has at least 10 standard stars.  The \emph{z} band was calibrated from the Landolt \emph{V}, \emph{B}, \emph{R} and \emph{I} band photometry, using the observed colour transformations given in table 7 of \citet{smi02}.

It was not possible to constrain the airmass coefficients from the standard star observations so average extinction factors from Subaru were used.  The colour coefficient, $B$, was poorly constrained in all bands, hence we have excluded this term in the calibration.

The zero-point values for each night were consistent, with final photometric errors of $< 0.05$ mag in all bands.  The photometric calibrations are summarised in table~\ref{tab:cal}.

\subsubsection{Optical, HST}

The optical ACS data was calibrated using the known calibrations provided in the fits image headers, and available from the Space Telescope Science Institute web-site\footnote{http://www.stsci.edu/hst/acs/analysis/zeropoints}.  The magnitudes are computed such that Vega would have zero magnitude in all bands.

\begin{table*}
\centering
\caption{Summary of photometric calibration.}
\label{tab:cal}
\begin{tabular}{lllll}
Waveband & $ZP$ & $A$ & $B$ & Colour for $B$ \\ \hline
\emph{K} &$22.37\pm 0.01$ & $-0.18 \pm 0.05$ & $0.00 \pm0.09$ & $J-K$ \\
\emph{J} &$22.78 \pm 0.01$& $-0.03 \pm 0.03 $& $-0.03 \pm 0.03$ & $J-K$ \\
\emph{z} & $27.08 \pm 0.02$ & 0 (fixed) & 0 &  --\\
\emph{I} &$26.26 \pm 0.04$& -0.03 (fixed) & 0 & --\\ 
\emph{R} &$27.50 \pm 0.03$& -0.09 (fixed)& 0& --\\
\emph{V} &$27.22 \pm 0.03$& -0.12 (fixed) & 0& --\\
\end{tabular}
\end{table*}

\subsection{Galaxies}
\label{galaxies}

Object detection was made in the \emph{K} band image.  The NIR is closely correlated with the dynamical mass of a galaxy (\citealt{gav96}), and therefore a good choice for selecting galaxies in an unbiased way for evolution studies.

The
{\sc SExtractor} software of \citet{ber96} was used to search for
objects in each field.  To detect objects a threshold value per pixel
must be chosen along with a minimum number of connected pixels.
Because the final \emph{K} band images are constructed from a jittered pattern of
images, the depth of observation varied across the final image, being
at its deepest in the centres and shallowest at the edges.  Therefore
detection of objects was done in two regions for each image, using
different detection parameters, in order to reach as deep as possible
in the centre of the image whilst avoiding spurious detections at the
edges of the image.  The minimum significance of the detections was $\approx 4\sigma$ in the centres of the images and $\approx 7\sigma$ at
the edges, where $\sigma$ is the background RMS determined from counts
over the whole image (and is thus an overestimate of $\sigma$ in the
central region and an underestimate in the outer region).  Faint pixels surrounding deblended objects are assigned to one of the sub-objects with a 
probability based on the expected contribution at that pixel from each of the deblended objects (see \protect\citealt{ber96}).  Reliability of the object detection, and in particular 
the handling of overlapping objects, was checked by eye. 

 Counts were measured using an adaptive aperture based
on Kron's algorithm (\citealt{kro80}), and also in a circular aperture with 
radius chosen to maximise the signal to noise.  For each object a value of
stellarity was also measured (see \citealt{ber96}).

To determine the magnitudes of the objects equation~\ref{eqn:mag} was used
with the following complication.  It is unknown to start with what the
true colours of the objects are, therefore the  last term in equation~\ref{eqn:mag} cannot
be determined.  To circumvent this problem an approximate magnitude was 
measured in each filter neglecting the colour term in equation~\ref{eqn:mag} allowing an approximate colour to be determined.  The approximate colour is then
multiplied by a correction factor previously determined using the same
technique to measure the approximate colours of standard stars and their
true colours.  The average colour correction for the NIR was
$\frac{(J-K)_{{\rm true}}}{(J-K)_{{\rm approx}}}=0.982$.


Note that for the determination of colours, magnitudes were derived from fixed,
circular apertures to avoid the effects of internal colour gradients, whereas the adaptive aperture magnitudes were 
used to derive pseudo-total magnitudes.  The use of adaptive aperture magnitudes
 using {\sc SExtractor}'s Kron radius should avoid the problem described by \citet{and02} of underestimating fluxes for galaxies with low central surface brightness. The overall reliability of the \emph{K} band photometry was checked
by comparing the field galaxy counts derived from the offset fields with deeper 
published 
results (see \citealt{ell04}). There is generally good agreement down
 to our limiting magnitudes.

 Because the
  \emph{J} band image has the same pixel scale as the \emph{K} band image it was possible 
to extract the galaxies at the positions measured in the \emph{K} band image using {\sc SExtractor}'s double image mode.  The galaxy colours were measured in the same size apertures as for the \emph{K} band data.

The \emph{V}, \emph{R}, \emph{I}, \emph{z}, \emph{F606W} and \emph{F814W} data covered a much larger area and had a different pixel scale and thus it was not possible to utilise the double image mode of {\sc SExtractor} to extract the galaxies.  Therefore catalogues of all the galaxies were made separately for each of the \emph{V}, \emph{R}, \emph{I}, \emph{z}, \emph{F606W} and \emph{F814W} band images using the same size apertures for the colours as for the NIR.
\emph{K} band selected galaxies were then extracted from this list using {\sc iraf}'s {\sc xyxymatch} task.  This takes the positions of three objects which are common to two images (preferably bright stars) and computes the transformation needed in terms of linear shifts, magnifications and rotations.  Thus a catalogue of \emph{K} selected galaxies can be collated in each band.

A few objects detected in the \emph{K} band were not detected in the optical bands due to either the faintness of the optical luminosity or the failure to deblend two very close \emph{K} band sources due to the different extents of the galaxy profiles in different bands.    
Note however, that the CMRs are determined only for galaxies with $K<19$ (except for fainter confirmed cluster members) and there is only one object not in \emph{F606W}, \emph{F814W}, \emph{R} and \emph{I} with $K<19$ and one more object in \emph{V}.  Both of these are due to the very small separation from another object and the failure
 to deblend and subsequently match these in the visible bands.  Thus at $K <19$ 
the matching of galaxies  is fairly robust and will not make a large difference to the derived CMRs.


Star--galaxy discrimination was determined using {\sc SExtractor's} stellarity parameter, in the \emph{K} band.  A cut-off of 0.8 was selected to delimit the two classes of objects with those objects with values greater than 0.8 being excluded as stars.  This value was confirmed by examination of the radial profiles of detected objects.  It
 was found that objects with a stellarity greater than 0.8 had an almost constant Gaussian FWHM close to the value of the seeing, whereas objects with stellarity less than 0.8 had more extended profiles.  The results are insensitive to the 
precise value of this cut-off as there were very few stars in each field.

\section{Morphological classification}
\label{sec:morph}

HST ACS data  have been used to morphologically classify the confirmed member galaxies.  These data have a point spread function (PSF) of $<0.1$ arcseconds and a pixel scale of 0.05 arcseconds.
Morphological classifications were carried out using 2-dimensional surface brightness fitting and isophotal classifications, as well as by inspection of the images by eye. We attemped to decompose the galaxies into bulge and disc components by
fitting a \citet{ser68} profile to the bulge and an exponential profile
to the disc following  \citet{kho00}. A pure elliptical
galaxy (no disc) is judged by a very small disc-to-total luminosity ratio
($D/T\lsim 0.05$) while a pure disc galaxy by $D/T \gtrsim 0.95$. The
quality of the fit is checked by the value of the reduced $\chi^{2}$ and visual
inspection of the residuals.  We do not rely on the 2D surface brightness
fit results if the reduced $\chi^{2}>2$. We use the HST PSF
simulator, {\sc Tiny Tim} (\citealt{kri95}) to simulate the PSF at the
position of each galaxy in the ACS image, although the fitting results are
found to be robust to the variations of the PSF across the camera. The images
in \emph{F814W} band are used for this analysis

In addition to the above method we also performed an isophotoal analysis
of the member galaxies via inspection of the radial surface brightness
and ellipticity profiles in two different bands (\emph{F814W} and \emph{F606W}). The final
categorisation  into early or late type galaxies has been decided by taking into account all the information from each technique on an individual basis.  All galaxies later than S0 were classified as late-types.     Generally galaxies which required a two component (bulge+disc) surface brightness profile, with the disc of equal or greater luminosity than the bulge were considered to be late types.  Bulge dominated systems were considered to be early-type galaxies unless there was a significant spiral or twisting structure in the isophotes.  Some galaxies (the BCG and two others) were not possible to classify into any conventional early or late type.

\section{Early type fraction}
\label{sec:earlyfrac}

Of our twenty-seven confirmed cluster members the fraction of early-type galaxies is 33 per cent (9 early-types; 15 late-types; 2 unclassified and the merging BCG).   This low value could be partly due to a selection effect, since it is easier to obtain redshifts for spectra of late-types due to prominent emission lines.  However, the galaxy selection for spectroscopy was made in the \emph{K} band which is insensitive to the presence of young stellar populations within the galaxies, and would therefore provide an unbiassed selection criterion, and focussed on red galaxies roughly within the colour-range of early-type galaxies at $z=0.89$.  
Furthermore,
selecting galaxies at brighter limiting magnitudes increases the spectroscopic completeness of the sample, and it is found that the early type fraction remains low in all cases.  Table~\ref{tab:earlyfrac} shows the early-type fractions at a range of limiting magnitudes.

\begin{table*}
\caption{The morphological fractions of confirmed cluster members.}
\label{tab:earlyfrac}
\begin{tabular}{lccccc}
Limiting & Spectroscopic & \multicolumn{3}{c}{Percentage of} &  Total \\
magnitude & completeness/ per cent & early-types & late-types & others & number \\ \hline
$K<19.6$ & 45 & 33 & 56 & 11 & 27\\
$K<19$ & 61 & 30 & 57 & 13 & 23\\
$K<18.5$ & 73 & 29 & 59 & 12 & 17\\
$K<18$ & 82 & 38 & 50 & 12 & 8\\
\end{tabular}
\end{table*}

\begin{figure}
\centering \includegraphics[scale=0.3,angle=270]{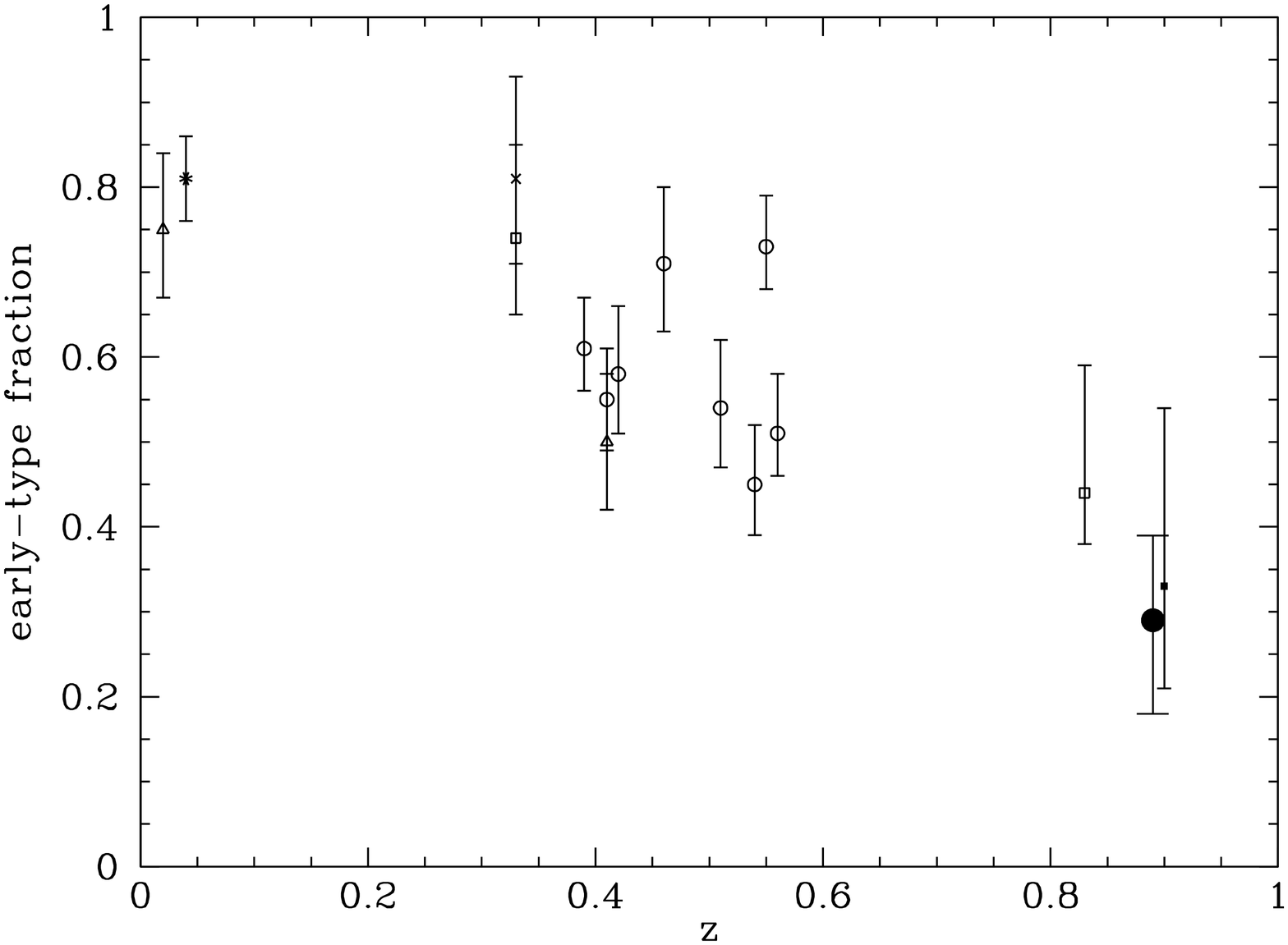}
\caption{A comparison of the early-type fraction in ClJ1226, indicated by the large closed circle, with those presented by \protect\citet{van00} (comprising of observations by \citealt{dre80}, star; \citealt{and97}, triangles; \citealt{dre97}, open circles; \citealt{lub98}, closed square; \citealt{fabr00}, cross; \citealt{van00}, open squares).}
\label{fig:earlyfrac}
\end{figure}

The low early-type fraction is indicative of a `morphological Butcher-Oemler effect', and agrees with previous results (e.g.\ \citealt{and97}; \citealt{dre97}; \citealt{cou98}; \citealt{van00}).  The early-type fraction as a function of redshift was presented by \citet{van00} and we show that the early-type fraction of ClJ1226 is in excellent agreement with this in Figure~\ref{fig:earlyfrac}.

\section{Colour-Magnitude Relations}
\label{sec:cmr}

CMRs were fit for three different sets: all galaxies excluding confirmed non-members, confirmed members, and confirmed early-type members.   Note that although the BCG shows classic signs of recent mergers, displaying multiple nuclei, it has been included in the early-type galaxies for the purpose of fitting the relations, since it clearly lies on the CMR in all cases and its extreme magnitude thus facilitates computing accurate slopes, normalisations and scatters for the CMRs.

\begin{figure*}
\begin{minipage}[c]{0.5\textwidth}
    \centering \includegraphics[scale=0.27,angle=270]
    {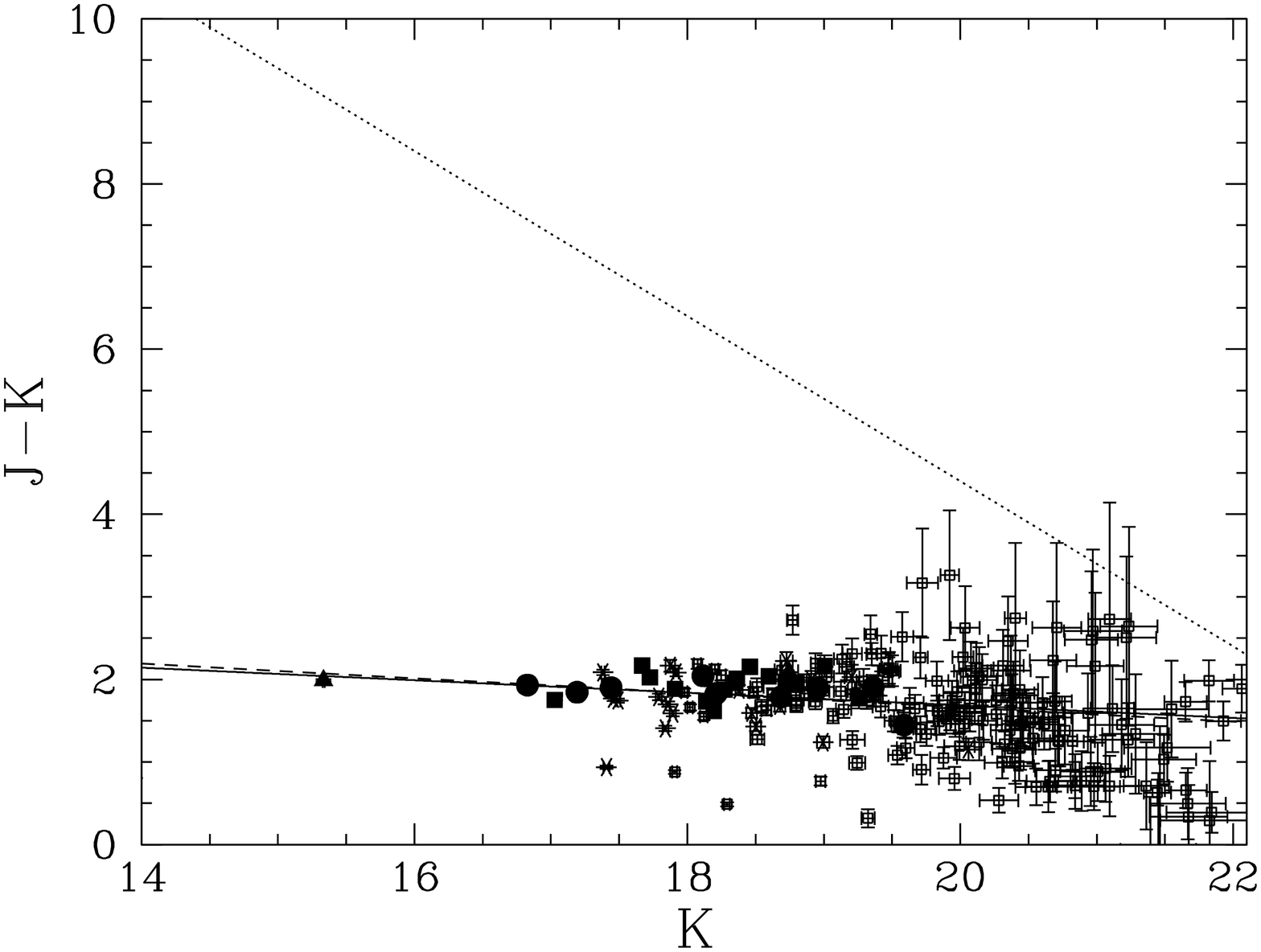}
  \end{minipage}%
  \begin{minipage}[c]{0.5\textwidth}
    \centering \includegraphics[scale=0.27,angle=270]
    {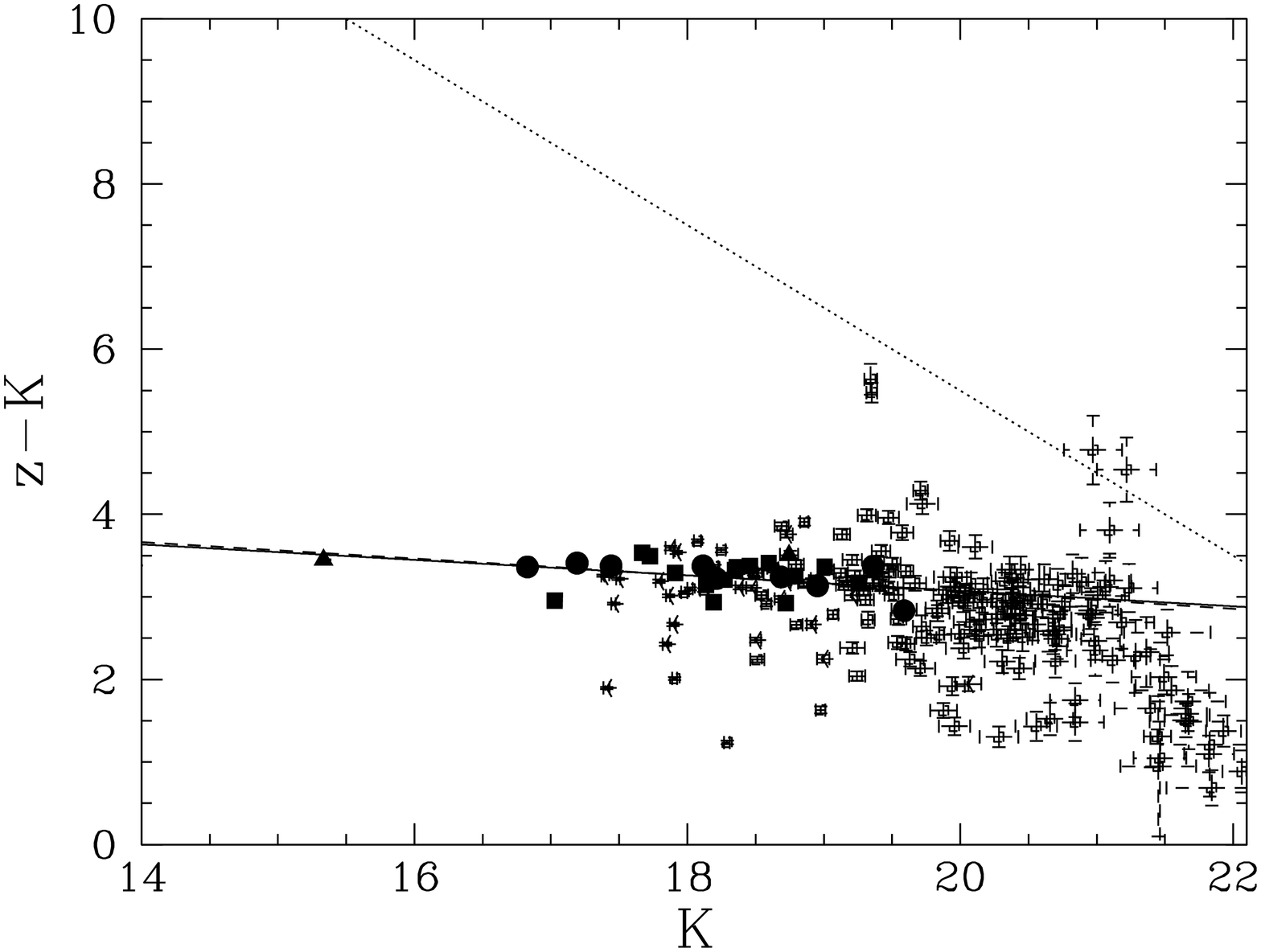}
  \end{minipage}
  \begin{minipage}[c]{0.5\textwidth}
    \centering \includegraphics[scale=0.27,angle=270]
    {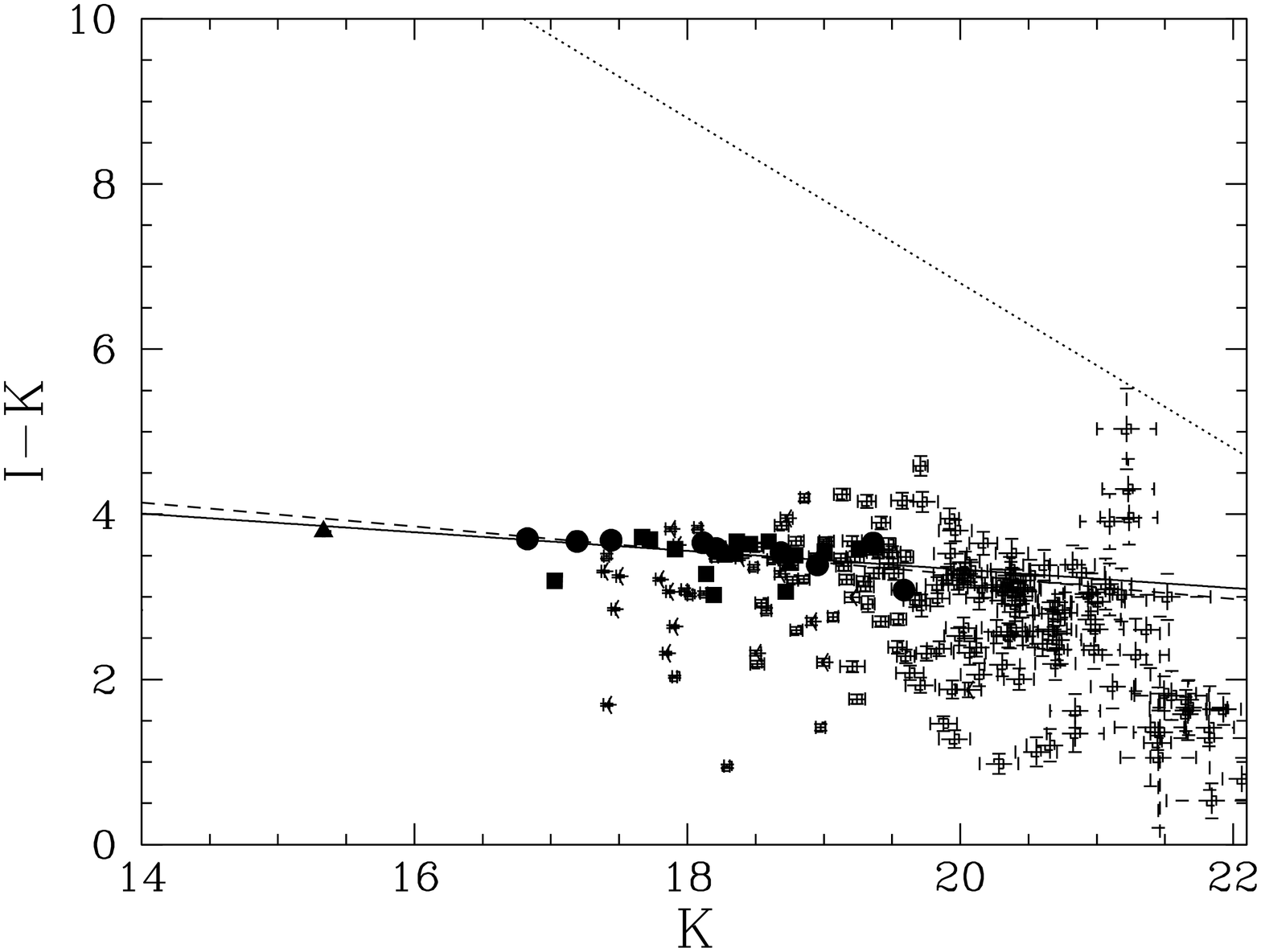}
  \end{minipage}%
  \begin{minipage}[c]{0.5\textwidth}
    \centering \includegraphics[scale=0.27,angle=270]
    {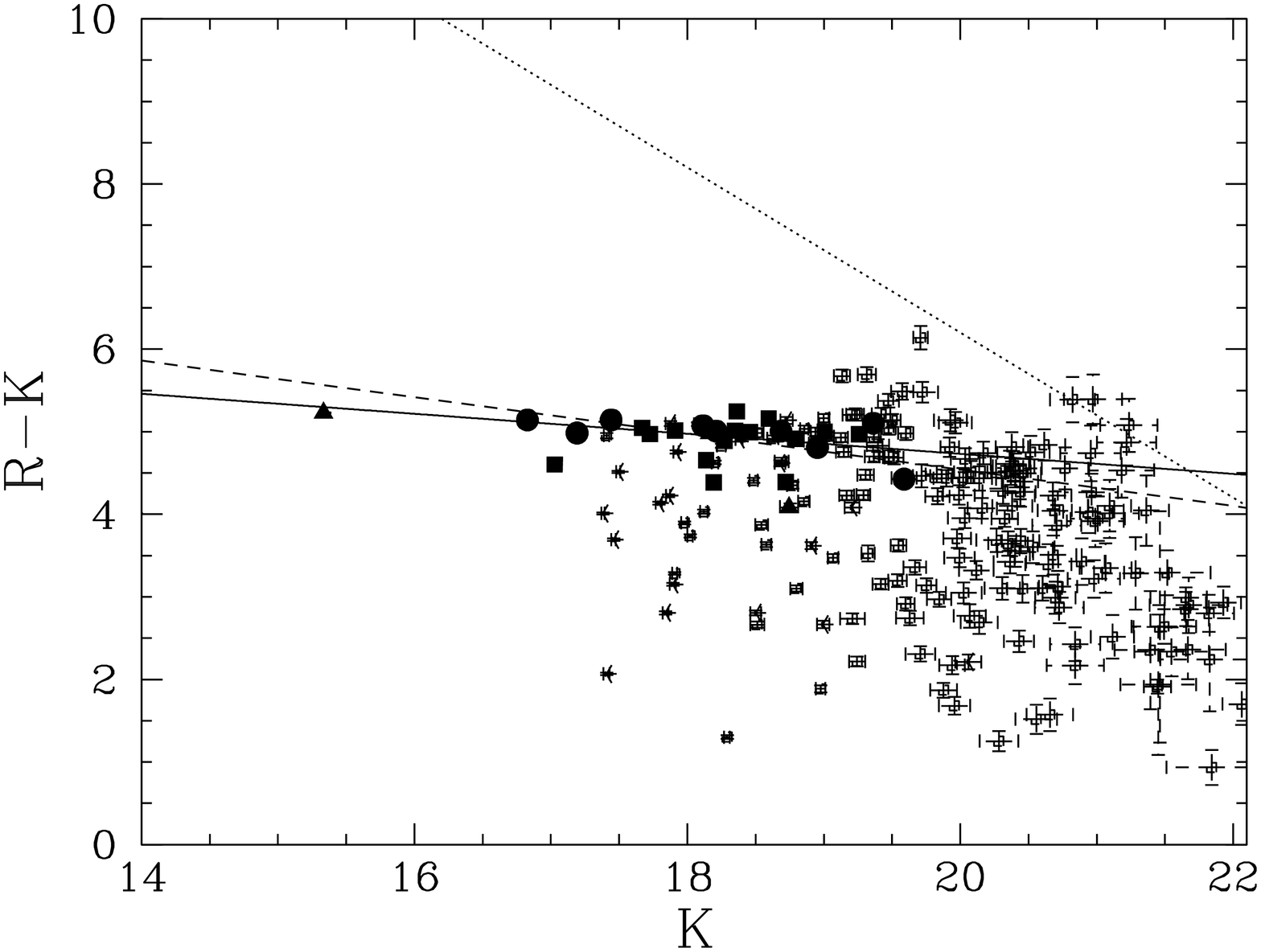}
  \end{minipage}
   \begin{minipage}[c]{0.5\textwidth}
    \centering \includegraphics[scale=0.27,angle=270]
    {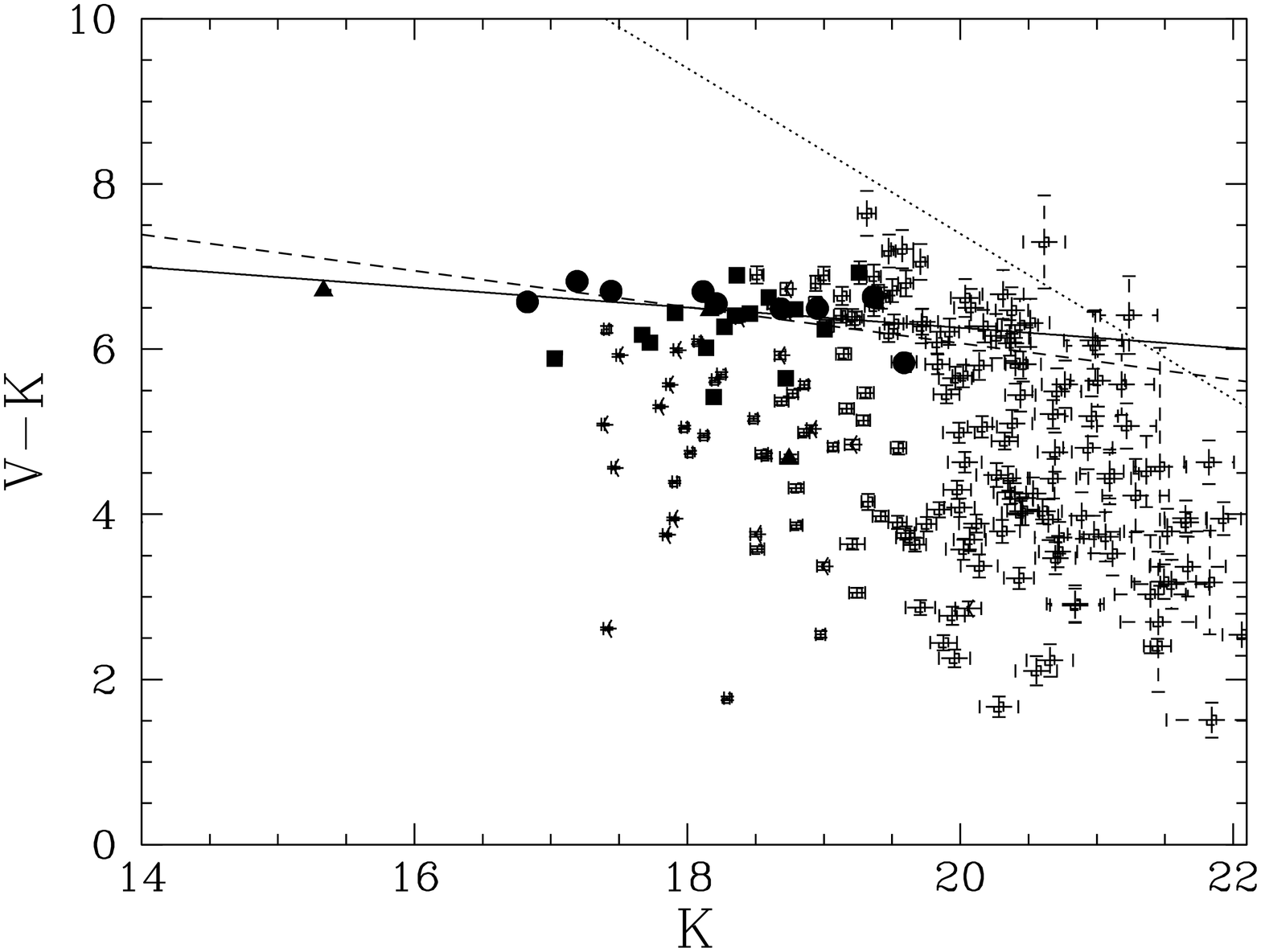}
  \end{minipage}%
  \begin{minipage}[c]{0.5\textwidth}
    \centering \includegraphics[scale=0.27,angle=270]
    {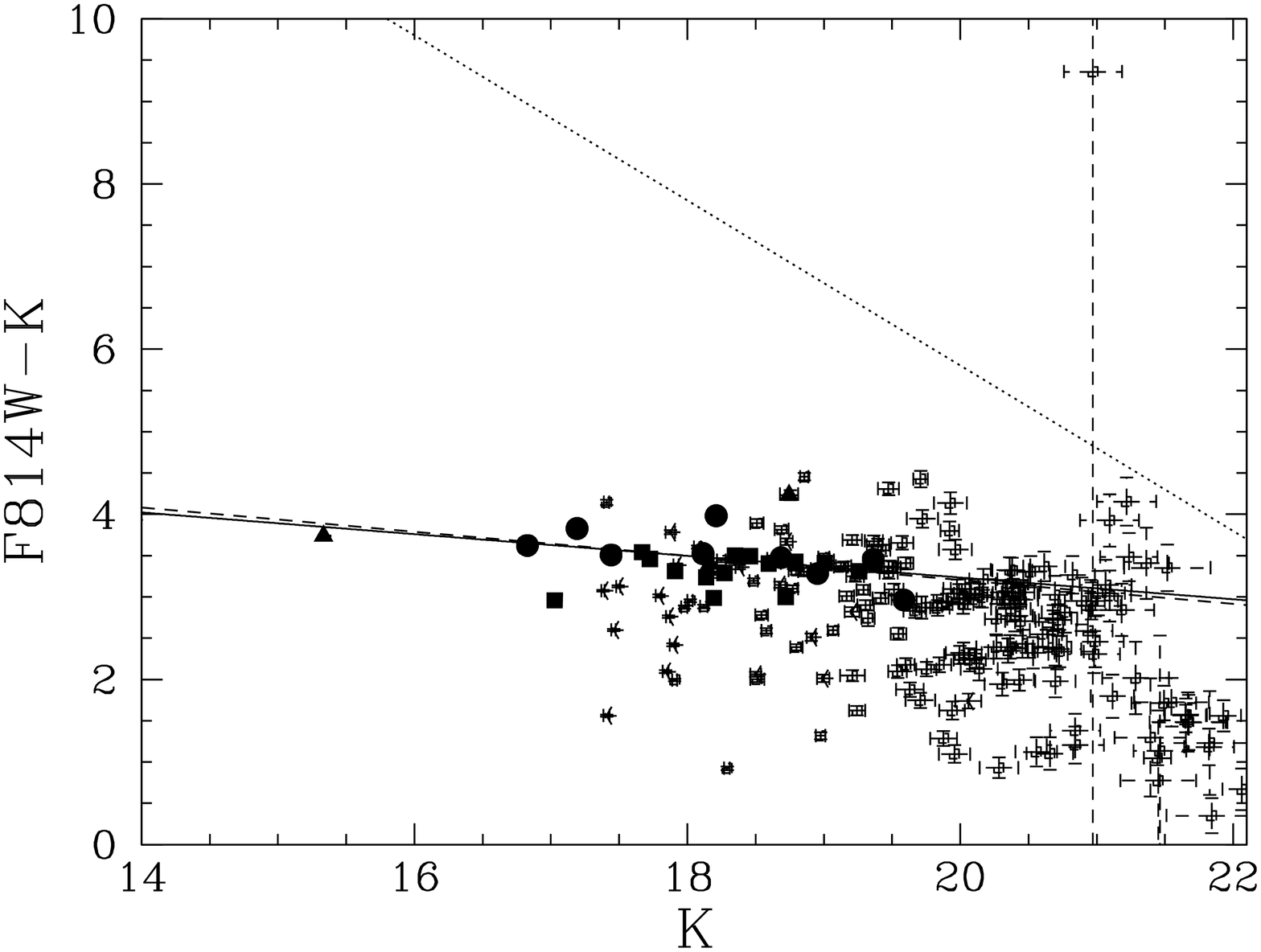}
  \end{minipage}
 \begin{minipage}[c]{0.5\textwidth}
    \centering \includegraphics[scale=0.27,angle=270]
    {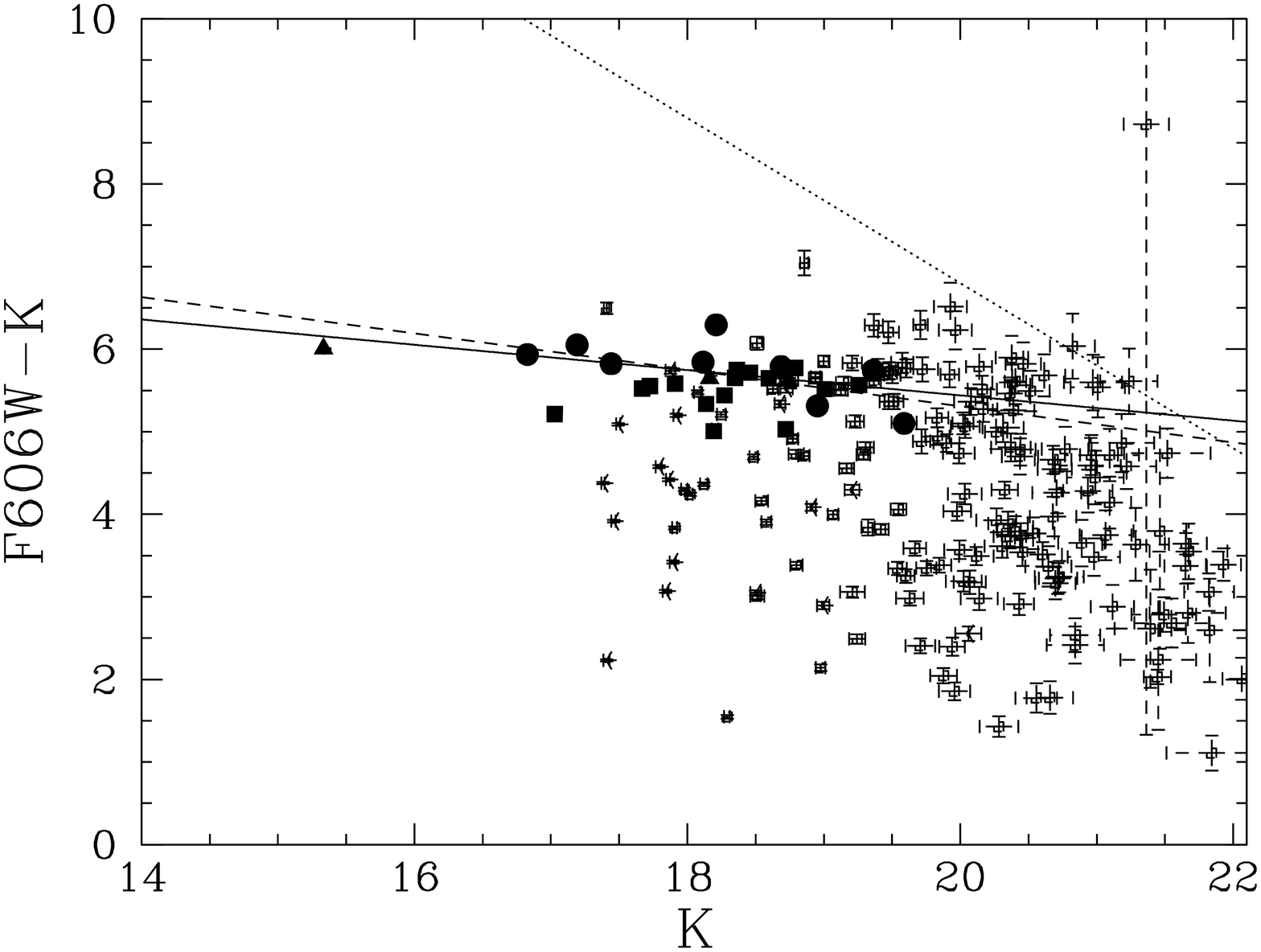}
  \end{minipage}

  \caption
    {Colour-magnitude relations for ClJ1226.  Filled circles indicate spectroscopically confirmed early-type cluster members, while the filled squares are the confirmed late-type members, filled triangles are cluster members with unclassified or unusual morphology, starred symbols indicate confirmed non-members, and open squares are of unknown membership.  The solid line shows the best fit to the early type galaxies and the BCG as discussed in the text.  The dashed line is the equivalent Coma relation normalised at $K^{*}=18.05$, see section~\ref{sec:slopes}.  The dotted line shows the approximate detection limit in each colour.}
  \label{fig:cmrs}
\end{figure*}

In order to fit the CMRs, the data were first divided into magnitude bins.  Each bin then contains a distribution of colours.  An estimate of the average value of this distribution was then estimated using the bi-weight estimator (\citealt{bee90}), where the average takes into account the 
skew of the distribution (due to a tight distribution of red galaxies associated with the cluster, and a population of bluer foreground galaxies).  Once the average for each magnitude bin was determined a straight line was fit using the method of least squares.  This straight line is the colour magnitude relation.   The results are given in Table~\ref{tab:slopes} and are plotted in Figure~\ref{fig:cmrs} along with the best fitting relations for the confirmed early-type member galaxies.  The diagonal dotted line shows the approximate detection limit; galaxies redder than this will not have been detected.  This was determined simply from the faintest reliable detection in each band.








\begin{table*}
\caption{The best fitting CMRs as fitted by the bi-weight estimator of \protect\citet{bee90}.  The CMRs have been fit separately for all galaxies (excluding confirmed non-members) with $K<19$ mag, confirmed members, and confirmed early-type members (plus the BCG).}
\label{tab:slopes}
\begin{tabular}{lcccccc}
& \multicolumn{2}{c}{All} & \multicolumn{2}{c}{Members} & \multicolumn{2}{c}{Early-types}\\
  & Intercept & Slope & Intercept & Slope & Intercept & Slope \\ \hline
$J-K$     &2.4 $\pm$ 0.3    & $-0.03 \pm 0.02$   & 3.1 $\pm 0.8$ & $-0.07 \pm 0.04$ & $3.2 \pm 0.7$ & $-0.08 \pm 0.04$ \\
$z-K$     & $4.32 \pm 0.09$ & $-0.057 \pm 0.005$ & $5.0 \pm 0.8$ & $-0.09 \pm 0.04$ & $4.9 \pm 0.8$ & $-0.09 \pm 0.04$\\
$I-K$     & $5.4 \pm 0.2$   & $-0.10 \pm 0.01$   & $5.7 \pm 0.7$ & $-0.12 \pm 0.04$ & $5.6 \pm 0.8$ & $-0.11 \pm 0.04$\\
$R-K$     & $7.5 \pm 0.5$   & $-0.14 \pm 0.03$   & $7.3 \pm 0.8$ & $-0.13 \pm 0.04$ & $7.2 \pm 0.9$ & $-0.12 \pm 0.05$\\
$V-K$     & $12 \pm 1$      & $-0.32 \pm 0.07$   & $9 \pm 1$     & $-0.12 \pm 0.07$ & $9 \pm 1$     & $-0.12 \pm 0.07$\\
$F\emph{606}W-K$ & $ 9.3 \pm 1.0 $ & $-0.21 \pm 0.06$   & $8.6 \pm 0.7$ & $-0.17 \pm 0.04$ & $9 \pm 1$     & $-0.15 \pm 0.07$\\
$F\emph{814W}-K$ & $5.8 \pm 0.6$   & $-0.14 \pm 0.04$   & $5.8 \pm 0.6$ & $-0.14 \pm 0.03$ & $5.9 \pm 1.0$ & $-0.13 \pm 0.05$\\
\end{tabular}
\end{table*}


The early-type member galaxies form an obvious red sequence in all cases.  However, some non early-type galaxies are also clearly on the CMR.  The BCG is a classic merging galaxy, showing double nuclei, and three very close companions, however it clearly lies on the CMR.  
Another intriguing result is the presence of late-type galaxies on the CMR.  We have conservatively adopted only E and S0 galaxies as early-types, all spirals are considered to be late-types.  The late-type galaxies on the CMR show a class of passive spirals, with the same photometric properties as elliptical galaxies, apparently consisting of similar populations of old stars with no obvious star-formation. 

\section{Evolution of the colour-magnitude relations}
\label{sec:evol}

The evolution of the CMR has been investigated via the scatter, the slope and the normalisation.  These investigations make extensive use of comparison to the nearby, $z=0.023$, Coma cluster, for which the data were kindly provided by Roberto de Propris and are taken from Eisenhardt et al. (2005, in preparation).  

 When comparing the CMR at $z=0.89$ with that at $z=0.023$ the shift in observed waveband must be accounted for.
Therefore in comparing the high $z$ CMRs with those of Coma two different relations are used.  The relations from the same
 filters are used to compare normalisations (section~\ref{sec:norm}) as the models used take into account the effects of redshifts and $k$-corrections.  However, for comparison of the slopes and scatters the nearest equivalent Coma relations are used, taking into account the shift in rest-frame wavebands when observing clusters at high redshifts.  The equivalent wavebands used are listed in Table~\ref{tab:scatter}.  Note that the at $z=0.89$ the \emph{V} band is equivalent to rest-frame wavelengths of $\approx 2700$\AA.  This is somewhat bluer than the \emph{U} band used as the nearest equivalent Coma waveband, and thus we emphasise that caution must be used in deriving conclusions from the comparison of the $V-K$ CMRs with the Coma $U-J$ CMRs.

\subsection{Scatter}
\label{sec:scatter}

The intrinsic scatter of the CMR is an extremely useful indicator of the star-formation and merger history of cluster galaxies.  However, to accurately quantify the intrinsic scatter it is first necessary to properly account for the contribution to the overall scatter introduced by statistical uncertainties inherent in measuring galaxy magnitudes.  We have followed the general method of \citet{sta98} to measure the intrinsic scatter and we give the details of the method below. 
We define the intrinsic scatter, $I_{{\rm S}}$ as,

\begin{equation}
\label{eq:scat}
I_{{\rm S}}=\sqrt{\sigma_{{\rm obs}}^{2} - \sigma_{{\rm phot}}^{2}},
\end{equation}

where $\sigma_{{\rm obs}}$ is the overall scatter of the CMR as observed and $\sigma_{{\rm phot}}$ is the scatter due to errors in the photometry.   The scatter and its error are determined using the biweight location and scale estimators (\citealt{bee90}) of the differences between the best fitting relation and the data.




It is paramount that $\sigma_{{\rm phot}}$ is determined accurately if the intrinsic scatter is to be meaningful.  We have followed \citet{sta98} in our method of determining $\sigma_{{\rm phot}}$ via Monte-Carlo simulations.  An artificial CMR was produced in which the magnitude of each galaxy is held at the value measured for each galaxy.  The colour of each galaxy was then set to a preliminary value determined from the best fitting relation.  This colour was then allowed to vary by a random amount determined from a pseudo-random Gaussian distribution approporiate to the measured error on the colour.  The scatter from this artificial relation was then calculated as before.  This process was repeated 1000 times to accurately quantify the photometric scatter, $\sigma_{{\rm phot}}$ and its error.

The intrinsic scatter was then calculated using equation~\ref{eq:scat}.  This process was repeated on three subsets of the \emph{K} band selected galaxies: all galaxies except confirmed non-members with $K<19$, all confirmed members, and all confirmed E and S0 galaxies plus the BCG.  The results are shown in Table~\ref{tab:scatter}.

\begin{table*}
\caption{The intrinsic scatter of the CMR, for all galaxies with $K<19$ mag except confirmed non-members, all confirmed members and all confirmed early-type members.  Also shown are the equivalent Coma relations for elliptical galaxies.}
\label{tab:scatter}
\begin{tabular}{lccclc}
& \multicolumn{3}{c}{Scatter} & Nearest equivalent & Scatter of\\
& All & Members & Early-type members & $z=0$ CMR & Coma ellipticals\\ \hline
$J-K$     & $0.3 \pm 0.4$ & $0.1 \pm 0.1$ & $0.1 \pm 0.1$   & $R-J$ & 0.05 $\pm$ 0.04\\
$z-K$     & $0.3 \pm 0.5$ & $0.1 \pm 0.1$ & $0.05 \pm 0.07$ & $V-J$ & $0.06 \pm 0.05$\\
$I-K$     & $0.4 \pm 0.6$ & $0.1 \pm 0.1$ & $0.05 \pm 0.04$ & $B-J$ & $0.06 \pm 0.04 $\\
$R-K$     & $0.6 \pm 0.8$ & $0.2 \pm 0.2$ & $0.08 \pm 0.05$ & $U-J$ & 0.09 $\pm$ 0.07\\
$V-K$     & 1.0 $\pm$ 0.7 & 0.2 $\pm$ 0.2 & 0.13 $\pm$ 0.10 & $U-J$ & 0.09 $\pm$ 0.07\\
$F\emph{606}W-K$ & $0.9 \pm 0.9$ & $0.2 \pm 0.1$ & $0.2 \pm 0.2$   & $U-J$ & 0.09 $\pm$ 0.07\\
$F\emph{814}W-K$ & $0.4 \pm 0.6$ & $0.1 \pm 0.1$ & $0.09 \pm 0.09$ & $B-J$ & $0.06 \pm 0.04 $
\end{tabular}
\end{table*}



Comparing the early-type relations to the Coma relations it can be seen that the high redshift relations are in good agreement with the nearest equivalent Coma relations.
These results support those found by \citet{sta98} for a larger number of clusters extending to similar redshifts, in which the scatter for all their clusters out to $z \approx 0.9$ is seen to be consistent with the equivalent Coma relations.  Similarly, \citet{ell97} find no increase in scatter out to $z \approx 0.54$.  Note that the inclusion of late-type galaxies increases the intrinsic scatter measured.

\subsection{Normalisation}
\label{sec:norm}

We have used the normalisation of the CMRs to measure the epoch of star-formation assuming an instantaneous burst of star-formation followed by passive evolution.  In theory it is possible to construct  more complex models.  In practice, however accurately modelling the star-formation histories of high redshift galaxies is well beyond the capabilities achievable through colour information of a single cluster alone. 
Therefore certain assumptions have been made to fix free parameters in the models. 



 
\subsubsection{Constructing the models}

\begin{figure}
\centering \includegraphics[width=0.75\columnwidth,angle=270]{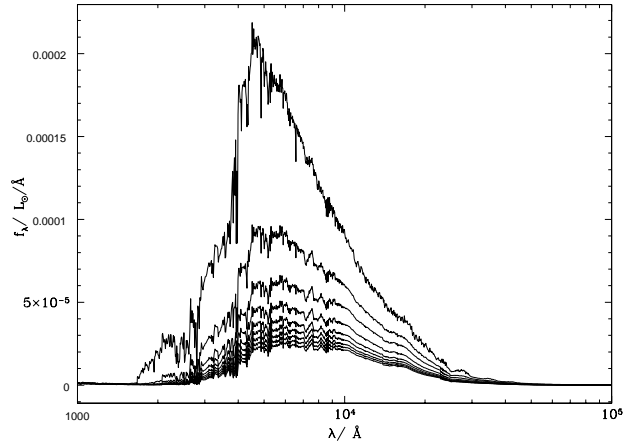}
\caption{Evolution of a passively evolving spectral energy distribution with Salpeter initial mass function and
 solar metallicity.  The spectra range from an age of 1 Gyr (brightest) to 9 Gyr (dimmest) in steps of 1 Gyr.  The spectra are displayed at a lower resolution than used in the modelling for clarity.}
\label{fig:spec_evol}
\end{figure}

\citet{bru03} provide libraries of synthetic stellar populations (SSPs) and their spectral evolution at unequal intervals of time from zero to twenty Gyr in the range $3200$\AA\ to $9500$\AA\ with a resolution of 3\AA.  We have followed the recommendation of using the Padova 1994 evolutionary tracks.  We assume the initial mass function (IMF) of \citet{cha03}.  Evolutionary tracks are provided with a range of metallicities denoted by $Z$, expressed as a fraction of the total mass, hence $Z_{\odot}=0.02$. 
 Figure~\ref{fig:spec_evol} shows an example of a passively evolving, solar metallicity spectral energy distribution evolves from an age of 1Gyr to 9Gyr, the youngest spectra are the brightest.

These SSPs assume an instantaneous burst of star-formation.  More elaborate star-formation histories may be modelled by combining multiple SSPs in an appropriate manner.   Here, however, we stick to the simple cases of purely passive evolution following an initial burst of star-formation at various redshifts, and a model in which no evolution occurs.

The spectra must be transformed to appear as they would to an observer.  A redshift of formation, $z_{{\rm f}}$, was chosen, corresponding to $t=0$ in the extracted spectra.  Age-redshift relations were calculated assuming a  cosmology with $H_{0}$=70 km s$^{-1}$ Mpc$^{-1}$, $\Omega_{{\rm M}}=0.3$ and $\Omega_{\Lambda}=0.7$.  Thus the redshift at which each extracted spectrum would be seen was calculated.  
Once the spectra were redshifted to appear as they would to an observer, they were convolved with the
 transmission function of the filter through which the observations to which the models will be compared were made.  
The spectra were then integrated yielding the flux per solar mass of the model galaxy.  The flux was then converted to magnitudes. assuming the magnitude of Vega is zero in all bands. The flux of Vega was calculated by integrating a calibrated model spectral energy distribution under the appropriate filter transmission.

Thus the evolution of the model galaxy spectra, expressed in terms of pseudo absolute magnitudes, was calculated at various wavelengths for redshifts of formation, $z_{{\rm f}}=1.5$, 3 and 5.  The magnitudes in different bands were then subtracted yielding the evolution in colour of the chosen model.  All the steps described above were achieved through use of code provided with the SSP libraries.




\subsubsection{Results}

\begin{figure*}
\begin{minipage}[c]{0.5\textwidth}
    \centering \includegraphics[scale=0.27,angle=270]
    {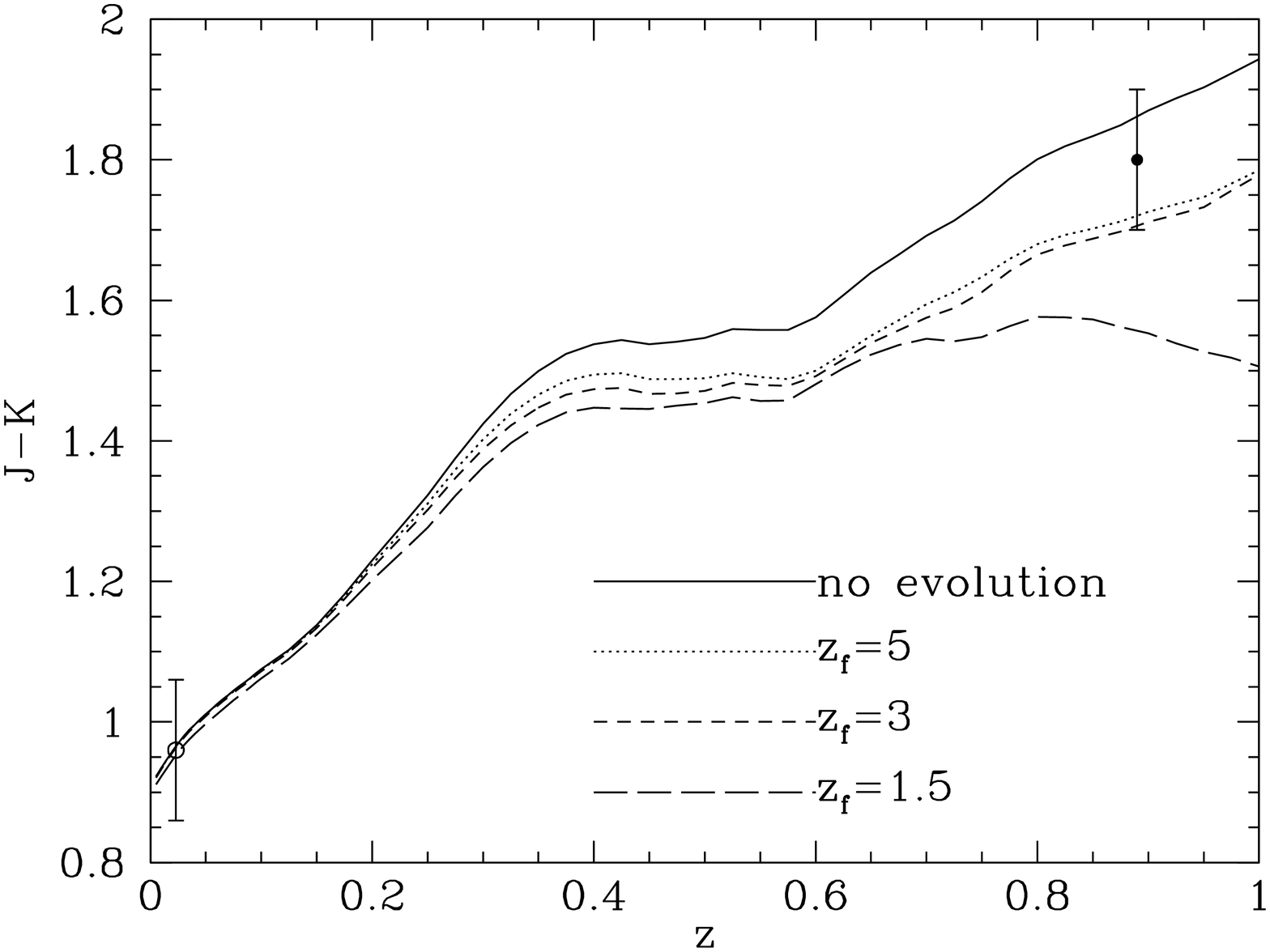}
  \end{minipage}%
  \begin{minipage}[c]{0.5\textwidth}
    \centering \includegraphics[scale=0.27,angle=270]
    {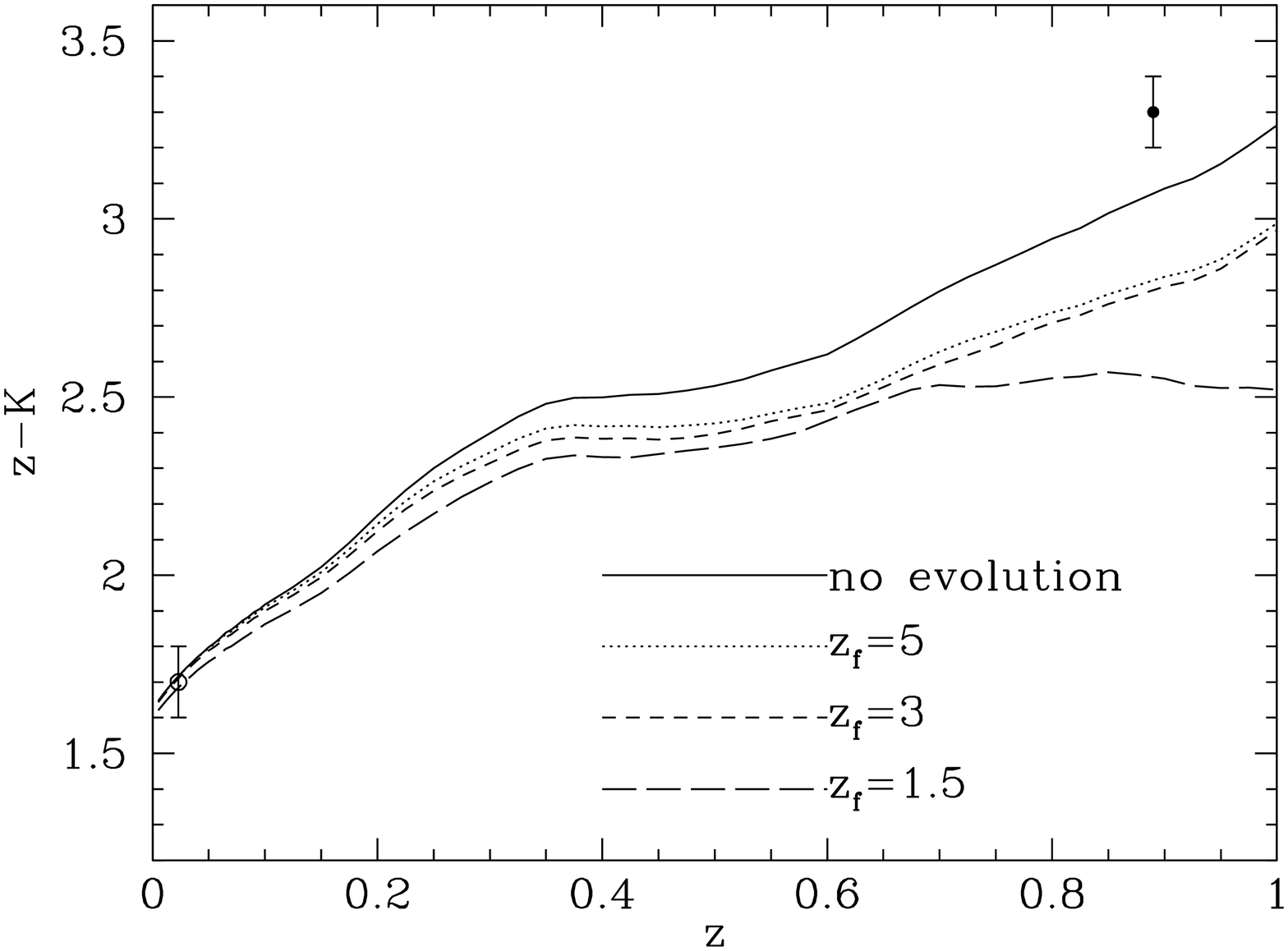}
  \end{minipage}
  \begin{minipage}[c]{0.5\textwidth}
    \centering \includegraphics[scale=0.27,angle=270]
    {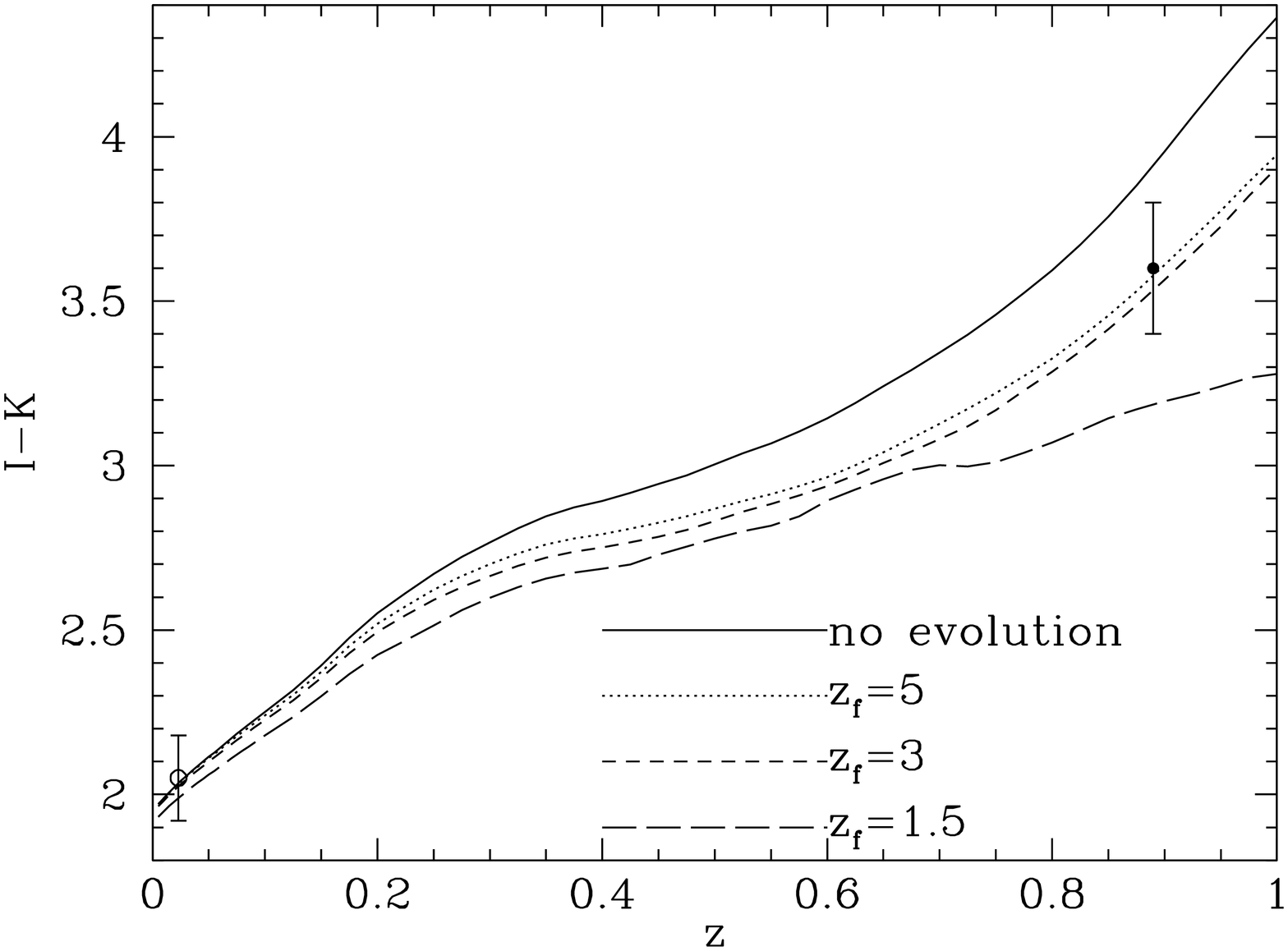}
  \end{minipage}%
  \begin{minipage}[c]{0.5\textwidth}
    \centering \includegraphics[scale=0.27,angle=270]
    {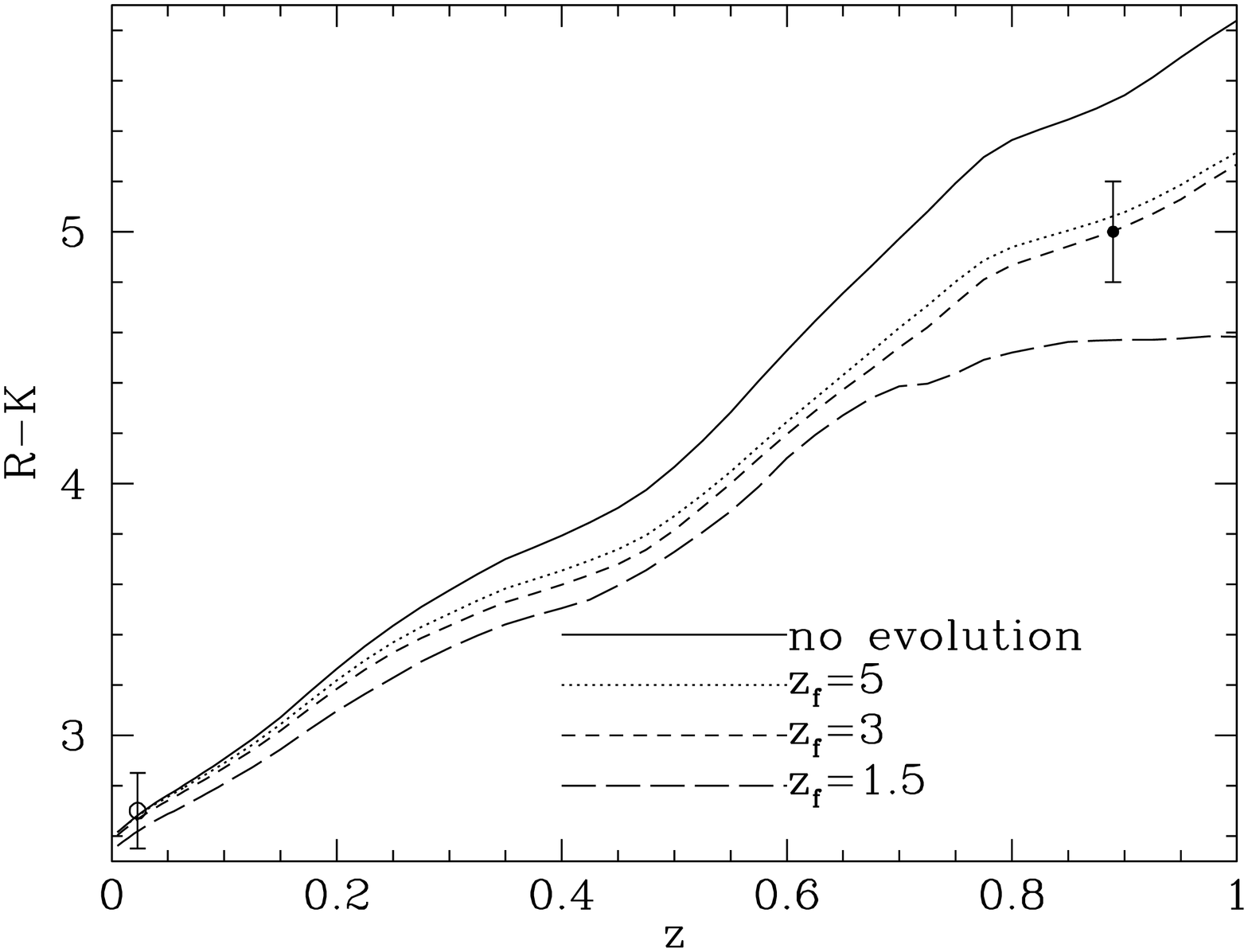}
  \end{minipage}
   \begin{minipage}[c]{0.5\textwidth}
    \centering \includegraphics[scale=0.27,angle=270]
    {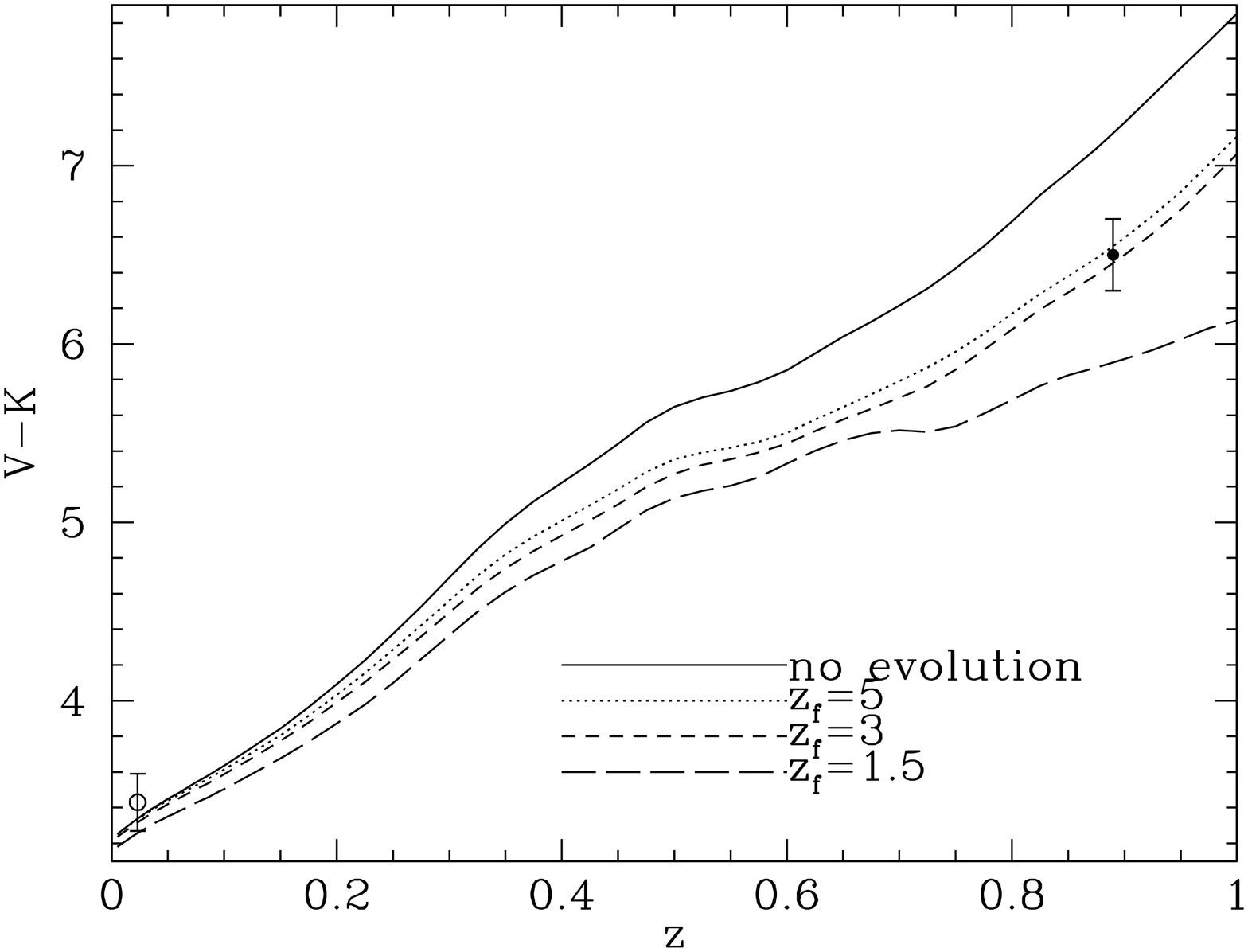}
  \end{minipage}%
  \begin{minipage}[c]{0.5\textwidth}
    \centering \includegraphics[scale=0.27,angle=270]
    {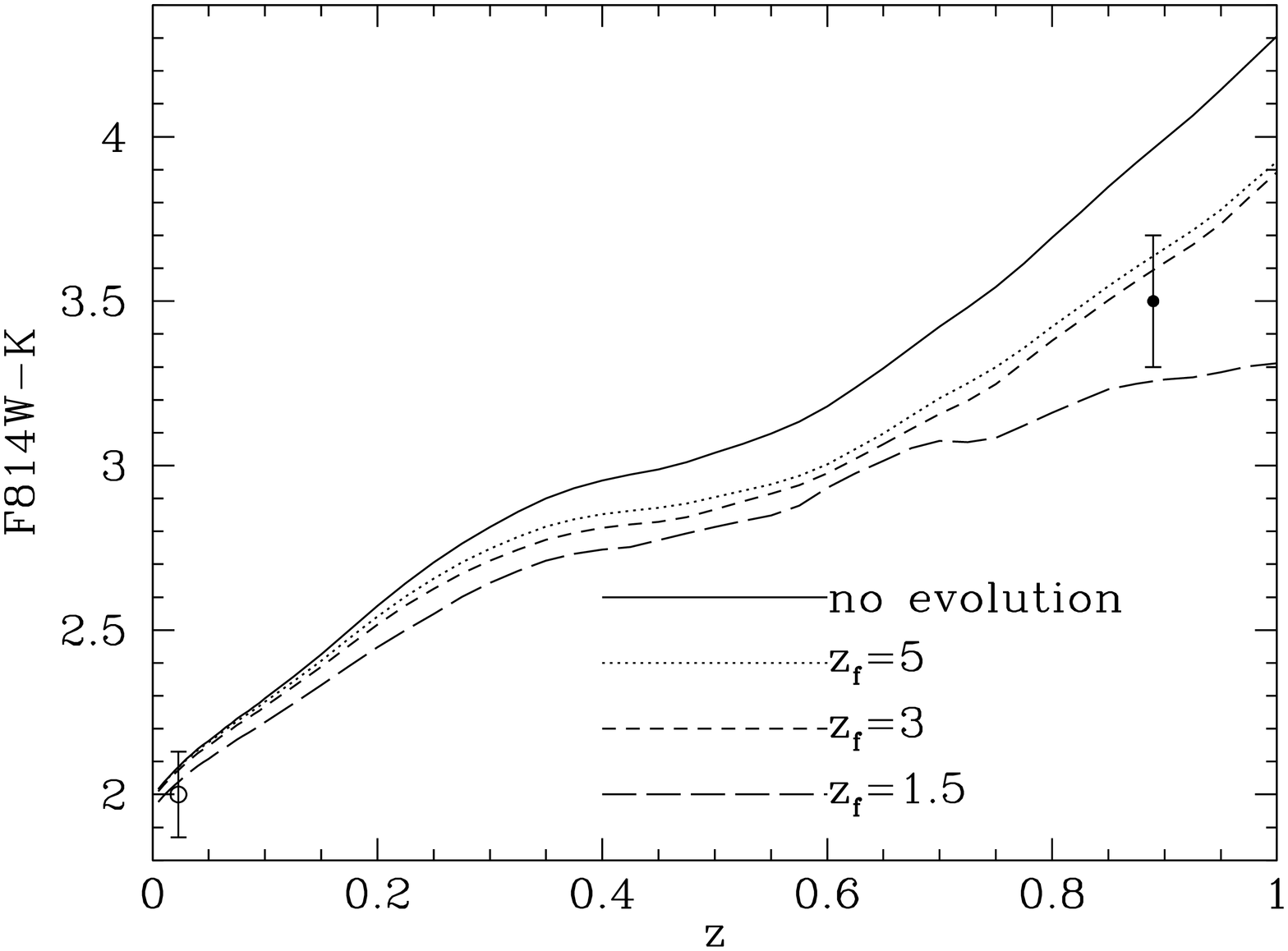}
  \end{minipage}
   \begin{minipage}[c]{0.5\textwidth}
    \centering \includegraphics[scale=0.27,angle=270]
    {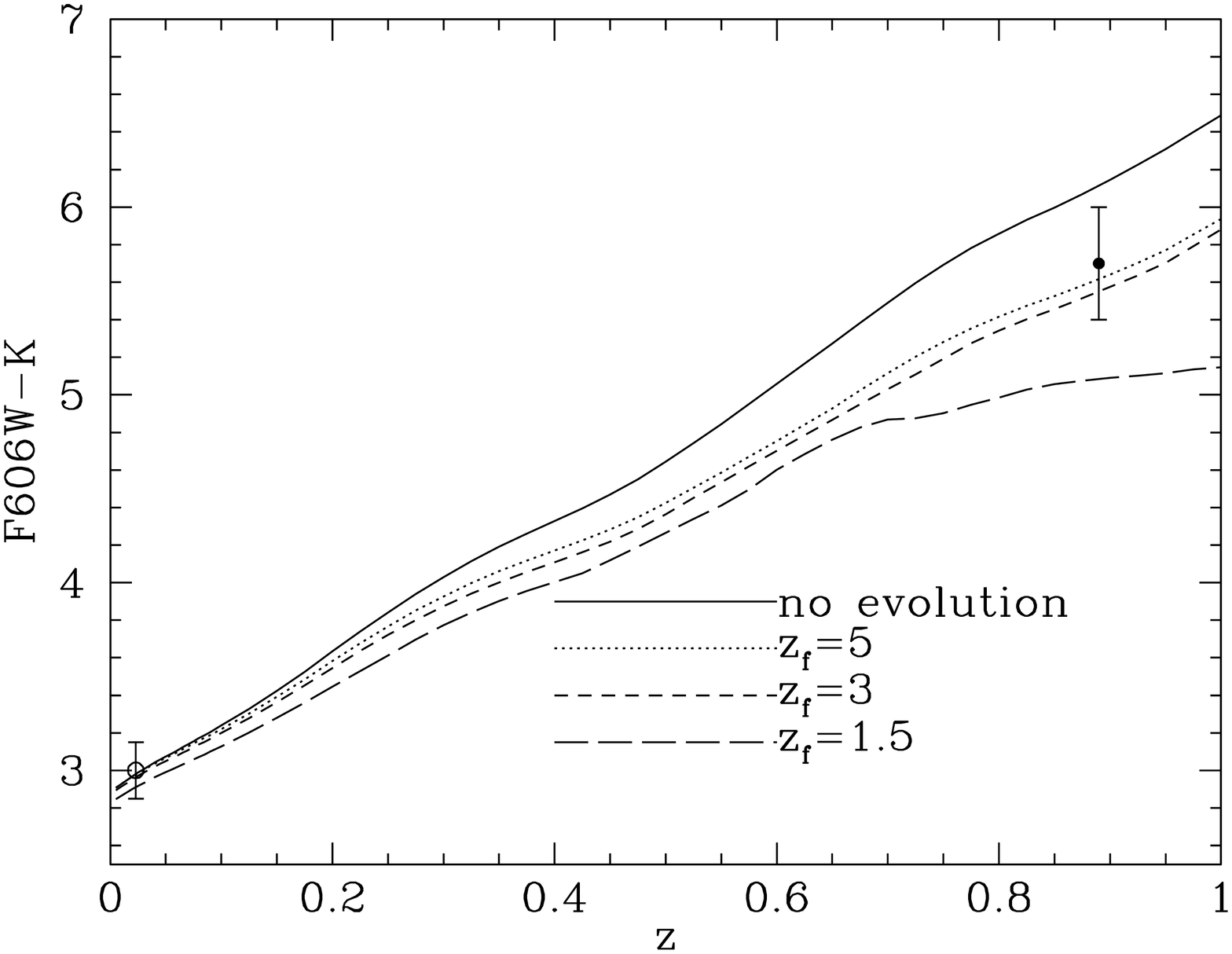}
  \end{minipage}%
    \caption
    {The normalisation of the CMR for ClJ1226 (closed circle) and Coma (open circle) compared to the evolution of models of a simple stellar population with $Z=Z_{\odot}$, Chabrier IMF, and $z_{{\rm f}}=$1.5, 3, 5 respectively.  Also shown is a no-evolution model.}
  \label{fig:ssp}
\end{figure*}

The models are shown in Figure~\ref{fig:ssp}.  In the observer's wavelength frame the colours become progressively bluer with age, since the galaxy is observed at smaller redshifts.  The evolving models display bluer colours than the no evolution model, in which the stellar populations do not evolve but are merely redshifted and k-corrected appropriately.  Furthermore, at fixed redshift younger stellar populations have bluer colours.

The models are compared to the value of the best fitting CMR evaluated at $K^{*}$ (ClJ1226, $K^{*}=18.05$, \citealt{ell04}; Coma, $K^{*}$=10.9, \citealt{dep98}) where the error on the ClJ1226 point is given by,

\begin{equation}
\sigma=\sqrt{\frac{\sum_{i=1}^{N}({\rm colour}_{i} - c - mK_{i})^{2}}{N-2}},
\end{equation}

where $c$ and $m$, are the intercept and slope of the best fitting relation respectively, and $N$ is the number of points.  The error on the Coma point is just the error on the intercept.  The Coma colours for the HST ACS filters were calculated using the ACS to Landolt system transformations given by \citet{sir05} for the E/S0 template at $z=0.023$.

At the low redshift of Coma the models are indistinguishable from each other and the Coma value can be achieved by all the models.  At $z=0.89$, however, the models diverge.  With the exception of the \emph{z} band all colours are consistent with formations at $z_{{\rm f}}\approx 3$ and $z_{{\rm f}}\approx 5$.  The \emph{z} band is not well fit by any relation, presumably due to systematic errors in the photometirc calibration.  This is very likely a  result of the transformations from the \emph{V}, \emph{B}, \emph{R} and \emph{I} band magnitudes of the standard stars of \citet{lan92} to the \emph{z} band, since the transformations used were derived from standard stars, and may not be applicable for the spectrum of galaxies at $z=0.89$.   Note, however, that though there may be a systematic offset in the zeropoint of the \emph{z} band, the measured scatter and slope of the CMR will not be affected by this.

\subsection{Slope}
\label{sec:slopes}

The slopes of the CMRs for ClJ1226 have been compared with those of Coma.  The slopes in each colour for ClJ1226 and those for the Coma relations and the nearest equivalent Coma relations (Eisenhardt et al. 2004, in preparation) are listed in Table~\ref{tab:comaslopes}.  Once again we 
point out that the equivalence between the relations is only approximate and for the $V-K$ relation there may be a significant difference between its true present day equivalent and the nearest equivalent, $U-J$, 
 relation used.  
The evolution of the slopes is illustrated in Figure~\ref{fig:cmrs}.  The figure shows the best fitting CMR compared to the equivalent Coma relations, which has been normalised at $K^{*}=18.05$.

\begin{table*}
\centering
\caption{Comparison of the best fitting colour-magnitude relations for ClJ1226 and Coma.}
\label{tab:comaslopes}
\begin{tabular}{lllll}
& ClJ1226 & \multicolumn{3}{l}{Coma}\\
&&Equivalent observed-frame &\multicolumn{2}{c}{Equivalent rest-frame}\\
Colour & Slope & slope & Colour &slope \\ \hline
$J-K$ & $-0.08 \pm 0.04$ &$-0.024 \pm 0.008$  & $R-J$ & $-0.088 \pm 0.008$\\
$z-K$ & $-0.09 \pm 0.04$ & $-0.069 \pm 0.011$ & $V-J$ & $-0.101 \pm 0.009$\\
$I-K$ & $-0.11 \pm 0.04$  &$-0.092 \pm 0.010$  & $B-J$ & $-0.146 \pm 0.010$\\
$R-K$ & $-0.12 \pm 0.05$ &$-0.113 \pm 0.012$  & $U-J$ & $-0.220 \pm 0.014$\\
$V-K$ & $-0.12 \pm 0.07$ &$-0.126 \pm 0.013$  & $U-J$ & $-0.220 \pm 0.014$\\
$F\emph{814}W-K$& $-0.13 \pm 0.05$ & $-0.092 \pm 0.010$ & $B-J$ & $-0.146 \pm 0.010$\\ 
$F\emph{606}W-K$& $-0.15 \pm 0.07$&$-0.113 \pm 0.012$  & $U-J$ & $-0.220 \pm 0.014$\\
\end{tabular}
\end{table*}

\citet{kod97} and \citet{kod98} present two models of evolution of the CMR, one in which the relation is driven by variation in the metallicities of galaxies of different magnitude, and one in which the sequence is driven by age.  Their models are capable of predicting the evolution of
 the slope of the CMR with redshift.  Figure 4 from \citet{kod98} shows that the expected evolution for a metallicity driven sequence is a slight steepening of the slope towards higher redshift, in the observed waveband, due largely to the blue-shifting of the rest-frame band-passes.  In the rest-frame the slope of the CMR will not change very much at all until ages younger than $\sim 4$Gyr, when it will become flatter due to the changes in metallicities (\citealt{kod97}, \citealt{kod98}, \citealt{gla98}).  There is little difference between their passive evolution, $z_{{\rm f}}=4.5$ model and their no evolution
 model.  A comparison of our best fitting slopes with that figure would seem to indicate the our results are consistent with their passively evolving metallicity sequence model, but inconsistent with their age-sequence model for which the slope is expected to evolve very rapidly with redshift. 
Essentially the same trends are also produced in the hierarchical models of \citet{kau98a} in which the largest elliptical galaxies are formed through the mergers of the largest, and therefore most metal rich, progenitor disc galaxies.




The slopes of the ClJ1226 CMRs  are generally in agreement, within errors, with the equivalent rest-frame Coma relations.  However, the ClJ1226 relations are systematically slightly flatter than the equivalent Coma relations.   Furthermore, the difference in the slopes is greater for bluer colours.  Note that here we are measuring something 
different from the expected steepening towards higher redshifts in the \citet{kod98} models, as those models compare low redshift and high redshift CMRs in the same observed colours, not the same rest-frame colours.

\section{Discussion} 
\label{sec:discuss}

Colour-magnitude relations of one of the most massive, high redshift clusters known have been determined.   The early-type galaxies comprise only a small fraction of the cluster, with only 33 per cent of the confirmed member galaxies having type E or S0.  This is indicative of a `morphological Butcher-Oemler effect', and is in good agreement with previous results (\citealt{and97}; \citealt{dre97}; \citealt{cou98}; \citealt{van00}).  The confirmed early-type members form a clear red-sequence in all colours.  The early-type galaxies display remarkably similar properties to early-type galaxies found in nearby clusters, suggesting that they are composed of similarly old stellar populations.

 The scatter observed is consistent with the  scatter in the nearest equivalent colours determined for elliptical galaxies in the Coma cluster within errors (Eisenhardt et al. 2004, in preparation), and suggests the stellar populations of early-type galaxies in ClJ1226 are already old at $z=0.89$.  This is in agreement with  previous work which has found no increase in scatter out to $z \approx 0.5$ (\protect\citealt{ell97}), $z \approx 0.9$ (\protect\citealt{sta98})  and even $z\approx 1.3$ (\citealt{bla03}, \citealt{hol04}).   \citet{bow98} discuss the effects of different star-formation histories on the scatter of the CMR.  The basic principle is that star-formation spread over a larger time would result in a larger scatter.  The colours also tend to approximately the same value $\gtrsim 5$ Gyr after star-formation ceases.  Thus the tight correlation of the CMR out to $z=0.89$ is highly suggestive that the majority of the star-formation occurred within a relatively short space of time or much earlier than the epoch at which the galaxies are observed.  \citet{bow98} also show that significant mergers would also disrupt the CMR causing an increase in scatter.  Even if most gas is already in the form of stars, in which case very little star-formation would be triggered by a merging event (\citealt{van01b}), there would be an increase in the scatter of the CMR due to the mixing of galaxy colours.  However, the effect is significantly smaller for for hierarchical merging as opposed to random merging (\citealt{bow98}).  \citet{kau98a}, show that the CMR can be reconstructed in a hierarchical scenario, if the progenitor galaxies have the necessary mass-metallicity relation.

The slopes of the CMRs are also consistent with those of the nearest equivalent Coma relations.  This is in agreement with the results of \citet{sta98} and \citet{hol04}. \citet{kod98} show that the evolution of the slope of the CMR in a sample of 17 clusters is consistent with the relation being a metallicity sequence as opposed to an age sequence.  Thus any processes which 
changes the metallicity of a galaxy would disrupt the slope.  A simple comparison of the measured values with the evolutionary tracks of \citet{kod98} 
suggest that our values are consistent with the $z_{{\rm f}}=4.5$, passively evolving metallicity sequence presented.  This is in agreement with \citet{gla98} who find an evolution of the slope of 6 clusters at $0.2 < z < 0.75$ and 44 clusters at $z <0.15$ is consistent with models of \citet{kod98}, although direct comparison of our results with \citet{gla98} is complicated by the different wavebands studied.
 
Our results are inconsistent with the findings of \citet{fer00}  who find an increase in slope and scatter of the UV$-$optical CMR of Cl 0939+4713 at $z=0.41$, which they argue is inconsistent with monolithic collapse.  They also point out the sensitivity of using the UV part of the spectrum as an indicator of star formation.  However, our results,  do analyse the restframe UV part of the spectrum since at $z=0.89$ \emph{V} is sampling about $2700$\AA.   The different results may be due to real differences in the star-formation histories of the different clusters examined.  Cl0939+4713 is thought to be part of a merging system (\citealt{schi96};  \citealt{def03}), with a large fraction of post-starburst galaxies (\citealt{bel95}), which may account for its enhanced star-formation.  However, XMM-Newton observations of ClJ1226 (Ben Maughan, private communication) also suggest that it is part of line-of-sight merger, but we do not see any evidence for enhanced star-formation.  On the other hand the different results may be in part due to differences in the analysis.   \citet{fer00} use a sample based on the morphological classifications of \citet{sma97}, which does not include redshift information, therefore the sample may be contaminated by foreground and background galaxies.   The detection limits of \citet{fer00} may also affect the steep slope they measure, since they may exclude galaxies on the predicted CMR fainter than $F\emph{702}W\approx 21.5$, but fit to galaxies down to $F\emph{702}W\approx 23.5$.  Note, however, this would make the scatter they measure even larger, though again we point out that the CMR may include non-cluster members.
Finally we note that we compute colours within a fixed aperture, whereas \citet{fer00} used the total magnitudes.

This issue of aperture size is important in deriving the colour-magnitude relation.
\citet{sco01} discusses the effect of using fixed aperture magnitudes compared to a fixed fraction of the galaxy light.  They find that the $U-V$ CMR of the Coma cluster is much flatter, and has a larger scatter when a fixed fraction of the galaxy light is used (a slope of $-0.016 \pm 0.018$ compared with $-0.074 \pm 0.008$).  They attribute these effects to the internal colour gradients of galaxies, which have a large scatter, and are generally redder towards the centre of elliptical galaxies.   Thus colours derived from fixed apertures will generally be redder for larger (and therefore usually brighter) galaxies.  However, the very tight relations of the Coma CMR (\citealt{ter01}) derived from fixed size apertures are difficult to understand in this scheme, since the large scatter of the colour gradients would also produce a large scatter at fixed radius.  An explanation may be that using a fixed aperture results in a higher signal to noise, and hence a tighter relation, or perhaps if the colour gradients were generally small, and hence the scatter introduced would be small.  Note that for the high-redshift galaxies reported here, the accurate computation of effective radius, and accurate photometry in large apertures is not possible, and thus a fixed angular aperture was used.  We investigated the effect of various aperture sizes on our results and found that for apertures larger than 1.5 arcseconds the measured colours were consistent. 

The effect of galaxy colour gradients could also explain the difference between the recent GALEX results from \citet{yi04}, which show  an increase in the scatter of the CMR when using near-ultraviolet$- $optical colours, derived from total magnitudes rather than aperture magnitudes.  However, the results of \citet{sco01} cannot explain the difference with \citet{fer00} who also use total magnitudes, but measure a steeper slope.

The merging model of \citet{kau98a} can reproduce the slope of the CMR at low redshifts due to the mass-metallicity relation of the progenitor galaxies being preserved in the hierarchical nature of the merging.  These models also predict a progressive flattening of the CMR at high redshift, becoming significant at $z>1$ due to  the younger average age of the most massive elliptical galaxies.  Some support for the merging model of \citet{kau98a} is found in the BCG of ClJ1226.  High resolution ACS data reveals that the BCG displays classic signs of merging activity: double nuclei and several close companions.  The BCG clearly lies on the CMR and has been included in measuring the evolution.  Such an obvious merger provides qualitative evidence that merging galaxies can remain on the CMR.  However we caution that BCGs are not typical galaxies, and very likely have distinct formation histories from the general cluster population (\citealt{bha85}).  \citet{bro02} and \citet{bro05} show that BCGs in clusters with high X-ray luminosity assembled most of their stellar mass at $z>1$ and thus the subsequent assembly of the BCG is probably not amongst equal mass galaxies, but is more likely a case of `galactic cannibalism' in which the BCG is much more massive than the galaxies with which it is merging.  Thus the merging activity may not have a large effect on the overall photometric properties of the BCG.  \citet{yam02} report a similar merging of the BCG in the $z=1.26$ cluster RXJ0848.9+4452, and \citet{van01a} show that the three brightest galaxies in the nearby cluster RXJ0848+4453 at $z=1.27$ have similarly undergone recent mergers.

On the other hand the merging BCG may be displaying behaviour typical of many early-type galaxies in clusters.  If mergers in clusters are predominantly between red galaxies, with old stellar population, as in MS1054-03 (\citealt{van99}), the colour-magnitude relation may be only a little affected during the mergers, displaying an increased scatter, and will quickly come to resemble the CMR of passive early-types (\citealt{tran05}).

The normalisation of the CMR shows evolution in the sense that galaxies at high redshift are generally bluer than their present day counterparts (with the exception of $z-K$, for which there may be significant systematic uncertainties in the photometric calibration, and $J-K$ which is consistent with no evolution within the errors), once the change in observed waveband has been taken into account.   This evolution is consistent with purely passive evolution in all cases except for $z-K$, with a redshift of formation $z_{{\rm f}} \gtrsim 3$.  These results are consistent with those of  \citet{sta98} and \citet{hol04}, who fit CMRs consistent with passive evolution. 

Thus the elliptical galaxies seem to be composed primarily of old stellar populations with very similar properties to those found in nearby systems.  This is consistent with the evolution of the \emph{K} band luminosity function (\citealt{ell04}) in which it is seen that the assembly history of the galaxies is consistent with monolithic collapse at $z_{{\rm f}} \gtrsim 2$.

An important point however, is that we have selected early-type galaxies to define our colour-magnitude relation.  Therefore we may be pre-selecting galaxies which contain only old stellar populations; the progenitor bias of \citet{van01b}.  Note if this is the case, then it follows that any galaxies either outside clusters at $z=0.89$, or of different morphological type, which are the progenitors of present day early-type galaxies in clusters, must end up with stellar populations indistinguishable from the early type galaxies already present in clusters at $z=0.89$, since the CMRs of clusters at high and low redshift are so similar. This could be the case if the progenitors are late-types containing similar populations of old-stellar populations, as well as younger populations.  Once star-formation ceases, the galaxy will quickly become redder, as the massive, hot, luminous, short-lived blue stars turn off the main sequence and enter the giant phase. Thus they could rather quickly resemble the old stellar populations of early-types soon after star-formation has ceased.  Of course at some point there must also be some morphological transformation.

The core ClJ1226 seems to be showing such evolution in  process.  There are four-late-type galaxies within 0.1 magnitudes of the best fitting CMR in $V-K$, which is the most discriminating CMR due to its large scatter.  These galaxies all possess substantial discs.  It would seem that these galaxies typify the process described above, whereby star-formation ceases in spirals and they migrate onto the CMR.  Their presence on the CMR is suggestive that evolution of the stellar-populations occurs before any morphological evolution which may take place.  This order of stellar evolution followed by any morphological evolution, is necessary to preserve the CMR at all redshifts, as observed by \citet{sta98}, otherwise the CMR would exhibit evolution in its scatter.

The presence of passive spirals have previously  been observed in high and intermediate redshift clusters  (e.g.\ \citealt{cou98}; \citealt{dre99}; \citealt{pog99}; \citealt{yam04}) and in low redshift clusters (\citealt{vandenbergh76}; \citealt{koop98}).  \citet{got03} observe that the passive spirals in these systems are found preferentially in the outskirts of clusters, where recently infalling galaxies are likely to be found, thereby providing a mechanism for the cessation of the star-formation, as the galaxies interact with the high-density environment.   Here however we find such passive galaxies in the core of an extremely massive, virialised cluster.  The presence of passive late-types in such a dense environment may be problematic if the transformation is entirely due to the infall of the galaxies.  Thus it may be that age is also important in the cessation of star-formation.  Galaxies in high density environments are likely to be older, since the high over-densities would collapse at earlier epochs in an expanding universe.  Therefore they could have already consumed their supply of gas by $z=0.89$, and be in the process of transformation.  Under such an explanation the morphology- and star-formation-density relation (eg.\ \citealt{lew02}) would be a reflection of an underlying age-density relation, due to the fact that galaxies form earlier on average in high-density environments.

Stronger constraints on the degree of morphological evolution and star-formation history should be possible via detailed spectroscopic and morphological studies of high redshift clusters as a function of density and environment.  The ACS data presented here will be further exploited in such studies to reveal finer details of the galaxy population of this remarkable cluster.

\section*{Acknowledgments}

We thank the referee for useful comments which have improved this paper.
The authors would like to thank Ben Maughan for providing his X-ray data, and useful discussions of the nature of ClJ1226, as well as his role in the observing.  Many thanks to Roberto de Propris for kindly providing the Coma data used extensively in this paper and to Adam Stanford for providing very useful help in the technical details of calculating the scatter of the CMR.  Thanks to Marco Sirianni for providing an early draft of his paper.
Thanks to the staff of UKIRT, Subaru, Keck, Gemini and STSCI observatories for providing wonderful facilities, and great support.
SCE wishes to acknowledge PPARC support.

\bibliographystyle{scemnras}
\bibliography{clusters}

\end{document}